\documentclass[twocolumn]{aastex62}

\graphicspath{{./}{figures/}}
\usepackage{enumitem}
\usepackage{epstopdf}

\shorttitle{Cepheids with \textit{TESS}: first light results}
\shortauthors{Plachy et al.}

\begin{document}

\title{\textit{TESS} observations of Cepheid stars: first light results}

\correspondingauthor{E. Plachy}
\email{eplachy@gmail.com, plachy.emese@csfk.mta.hu}

\author[0000-0002-5481-3352]{E. Plachy}

\affiliation{Konkoly Observatory, Research Centre for Astronomy and Earth Sciences, MTA Centre of Excellence, Konkoly Thege Mikl\'os \'ut 15-17, H-1121 Budapest, Hungary}
\affiliation{MTA CSFK Lend\"ulet Near-Field Cosmology Research Group}
\affiliation{ELTE E\"otv\"os Loránd University, Institute of Physics, 1117, P\'azm\'any P\'eter s\'et\'any 1/A, Budapest, Hungary}

\author[0000-0001-5449-2467]{A. P\'al}
\affiliation{Konkoly Observatory, Research Centre for Astronomy and Earth Sciences, MTA Centre of Excellence, Konkoly Thege Mikl\'os \'ut 15-17, H-1121 Budapest, Hungary}
\affiliation{ELTE E\"otv\"os Loránd University, Institute of Physics, 1117, P\'azm\'any P\'eter s\'et\'any 1/A, Budapest, Hungary}

\author[0000-0002-8585-4544]{A. B\'odi}
\affiliation{Konkoly Observatory, Research Centre for Astronomy and Earth Sciences, MTA Centre of Excellence, Konkoly Thege Mikl\'os \'ut 15-17, H-1121 Budapest, Hungary}
\affiliation{MTA CSFK Lend\"ulet Near-Field Cosmology Research Group}
\affiliation{ELTE E\"otv\"os Loránd University, Institute of Physics, 1117, P\'azm\'any P\'eter s\'et\'any 1/A, Budapest, Hungary}

\author[0000-0002-5781-1926]{P. Szab\'o}
\affiliation{Konkoly Observatory, Research Centre for Astronomy and Earth Sciences, MTA Centre of Excellence, Konkoly Thege Mikl\'os \'ut 15-17, H-1121 Budapest, Hungary}
\affiliation{MTA CSFK Lend\"ulet Near-Field Cosmology Research Group}
\affiliation{Magdalene College, University of Cambridge, CB3 0AG, Cambridge, UK}

\author[0000-0002-8159-1599]{L. Moln\'ar}
\affiliation{Konkoly Observatory, Research Centre for Astronomy and Earth Sciences, MTA Centre of Excellence, Konkoly Thege Mikl\'os \'ut 15-17, H-1121 Budapest, Hungary}
\affiliation{MTA CSFK Lend\"ulet Near-Field Cosmology Research Group}
\affiliation{ELTE E\"otv\"os Loránd University, Institute of Physics, 1117, P\'azm\'any P\'eter s\'et\'any 1/A, Budapest, Hungary}

\author[0000-0002-2046-4131]{L. Szabados}
\affiliation{Konkoly Observatory, Research Centre for Astronomy and Earth Sciences, MTA Centre of Excellence, Konkoly Thege Mikl\'os \'ut 15-17, H-1121 Budapest, Hungary}
\affiliation{MTA CSFK Lend\"ulet Near-Field Cosmology Research Group}

\author[0000-0003-3851-6603]{J.~M. Benk\H{o}}
\affiliation{Konkoly Observatory, Research Centre for Astronomy and Earth Sciences, MTA Centre of Excellence, Konkoly Thege Mikl\'os \'ut 15-17, H-1121 Budapest, Hungary}
\affiliation{MTA CSFK Lend\"ulet Near-Field Cosmology Research Group}

\author[0000-0001-8089-4419]{R. I. Anderson}
 \affiliation{European Southern Observatory, Karl-Schwarzschild-Str. 2, 85748 Garching b. M\"{u}nchen, Germany}
 
\author[0000-0003-4456-4863]{E. P. Bellinger}
\affiliation{Stellar Astrophysics Centre, Department of Physics and Astronomy, Aarhus University, Ny Munkegade 120, DK-8000 Aarhus C, Denmark}

\author[0000-0001-6147-3360]{A. Bhardwaj}
\affiliation{Korea Astronomy and Space Science Institute, Daedeokdae-ro 776, Yuseong-gu, Daejeon 34055, Republic of Korea}

\author{M. Ebadi}
\affiliation{Solar Physics and Astronomy Section, Institute of Geophysics, University of Tehran, Iran}
\author[0000-0002-8855-3923]{K. Gazeas}

\affiliation{Section of Astrophysics, Astronomy and Mechanics, Department of Physics, National and Kapodistrian University of Athens, GR-15784
Zografos, Athens, Greece}

\author{F.-J. Hambsch}
\affiliation{Vereniging Voor Sterrenkunde (VVS), Oostmeers 122 C, 8000 Brugge, Belgium}
\affiliation{Bundesdeutsche Arbeitsgemeinschaft f\"ur Ver\"anderliche Sterne e.V. (BAV), Munsterdamm 90, D-12169 Berlin, Germany}
\affiliation{American Association of Variable Star Observers, 49 Bay State Road, Cambridge, MA 02138, USA}

\author[0000-0002-7286-1438]{A. Hasanzadeh} 
\affiliation{Solar Physics and Astronomy Section, Institute of Geophysics, University of Tehran, Iran}

\author[0000-0002-8591-4295]{M. I. Jurkovic} 
\affiliation{Astronomical Observatory of Belgrade, Volgina 7., 11 060 Belgrade, Serbia}
\affiliation{Konkoly Observatory, Research Centre for Astronomy and Earth Sciences, MTA Centre of Excellence, Konkoly Thege Mikl\'os \'ut 15-17, H-1121 Budapest, Hungary}

\author{M. J. Kalaee}
\affiliation{Solar Physics and Astronomy Section, Institute of Geophysics, University of Tehran, Iran}
\affiliation{Space Physics Group, Institute of Geophysics, University of Tehran, Iran}

\author[0000-0003-0626-1749]{P. Kervella}
\affiliation{LESIA, Observatoire de Paris, Universit\'e PSL, CNRS, Sorbonne Universit\'e, Universit\'e de Paris, 5 Place Jules Janssen, 92195
Meudon, France}

\author{K. Kolenberg}
 \affiliation{Institute of Astronomy, KU Leuven, Celestijnenlaan 200D, B-3001 Heverlee, Belgium}
 \affiliation{Physics Department, University of Antwerp, Groenenborgerlaan 171, B-2020
Antwerpen, Belgium}
\affiliation{Department of Physics and Astronomy, Vrije Universiteit Brussel - VUB, Pleinlaan 2, B-1050 Brussel, Belgium}

\author[0000-0001-8916-8050]{P. Miko{\l}ajczyk}
\affiliation{Astronomical Institute, University of Wroc{\l}aw, ul. Kopernika 11, 51-622
Wroc{\l}aw, Poland}

\author[0000-0002-7399-0231]{N. Nardetto}
 \affiliation{
 Universit\'e C\^ote d'Azur, Observatoire de la C\^ote d'Azur, CNRS, Laboratoire Lagrange, France}
\author{J. M. Nemec}
\affiliation{Department of Physics \& Astronomy, Camosun College, Victoria, BC, Canada}

\author{H. Netzel}
\affiliation{Nicolaus Copernicus Astronomical Center, Polish Academy of Sciences, PAS, Bartycka 18, PL-00-716 Warsaw, Poland}

\author[0000-0001-8771-7554]{C.--C. Ngeow}
\affiliation{Graduate Institute of Astronomy, National Central University, Jhongli 32001, Taiwan}

\author[0000-0001-8544-0950]{D. Ozuyar}
\affiliation{Ankara University, Faculty of Science, Dept. of Astronomy and Space Sciences, 06100, Tandogan, Ankara, Turkey}

\author[0000-0003-0139-6951]{J. Pascual-Granado}
\affiliation{Instituto de Astrof\'isica de Andaluc\'ia (CSIC), Glorieta de la Astronom\'ia
s/n, E-18008 Granada, Spain}

\author[0000-0003-3861-8124]{B. Pilecki} 
\affiliation{Nicolaus Copernicus Astronomical Center, Polish Academy of Sciences, PAS, Bartycka 18, PL-00-716 Warsaw, Poland}

\author{V. Ripepi}
\affiliation{INAF-Osservatorio Astronomico di Capodimonte, via Moiariello 16, I-80131 Naples, Italy}
\author[0000-0002-7602-0046]{M. Skarka}
\affiliation{Department of Theoretical Physics and Astrophysics, Masaryk University, Kotl\'{a}\v{r}sk\'{a} 2, 61137 Brno, Czech Republic}
\affiliation{Astronomical Institute, Czech Academy of Sciences, Fri\v{c}ova 298, 25165, Ond\v{r}ejov,
Czech Republic}

\author[0000-0001-7217-4884]{R. Smolec}
\affiliation{Nicolaus Copernicus Astronomical Center, Polish Academy of Sciences, PAS, Bartycka 18, PL-00-716 Warsaw, Poland}

\author[0000-0001-7806-2883]{\'A. S\'odor}
\affiliation{Konkoly Observatory, Research Centre for Astronomy and Earth Sciences, MTA Centre of Excellence, Konkoly Thege Mikl\'os \'ut 15-17, H-1121 Budapest, Hungary}
\affiliation{MTA CSFK Lend\"ulet Near-Field Cosmology Research Group}

\author[0000-0002-3258-1909]{R. Szab\'o}
\affiliation{Konkoly Observatory, Research Centre for Astronomy and Earth Sciences, MTA Centre of Excellence, Konkoly Thege Mikl\'os \'ut 15-17, H-1121 Budapest, Hungary}
\affiliation{MTA CSFK Lend\"ulet Near-Field Cosmology Research Group}
\affiliation{ELTE E\"otv\"os Loránd University, Institute of Physics, 1117, P\'azm\'any P\'eter s\'et\'any 1/A, Budapest, Hungary}

\author{J. Christensen-Dalsgaard}
\affiliation{Stellar Astrophysics Centre, Department of Physics and Astronomy, Aarhus University, Ny Munkegade 120, DK-8000 Aarhus C, Denmark}

\author[0000-0002-4715-9460]{J. M. Jenkins}
\affiliation{NASA Ames Research Center, Moffett Field, CA 94035, USA}

\author[0000-0002-9037-0018]{H. Kjeldsen}
\affiliation{Stellar Astrophysics Centre, Department of Physics and Astronomy, Aarhus University, Ny Munkegade 120, DK-8000 Aarhus C, Denmark}

\author{G. R. Ricker}
\affiliation{Department of Physics and Kavli Institute for Astrophysics and Space Research, Massachusetts Institute of Technology, Cambridge, MA 02139, USA}

\author[0000-0001-6763-6562]{R. Vanderspek}
\affiliation{Department of Physics and Kavli Institute for Astrophysics and Space Research, Massachusetts Institute of Technology, Cambridge, MA 02139, USA}


\begin{abstract}


We present the first analysis of Cepheid stars observed by the \textit{TESS} space mission in Sectors 1 to 5. Our sample consists of 25 pulsators: ten fundamental mode, three overtone and two double-mode classical Cepheids, plus three Type~II and seven anomalous Cepheids. The targets were chosen from fields with different stellar densities, both from the Galactic field and from the Magellanic System. Three targets have 2-minute cadence light curves available by the TESS Science Processing Operations Center: for the rest, we prepared custom light curves from the full-frame images with our own differential photometric FITSH pipeline. Our main goal was to explore the potential and the limitations of \textit{TESS} concerning the various subtypes of Cepheids. We detected many low amplitude features: weak modulation, period jitter, and timing variations due to light-time effect. We also report signs of non-radial modes and the first discovery of such a mode in an anomalous Cepheid, the overtone star XZ Cet, which we then confirmed with ground-based multicolor photometric measurements. We prepared a custom photometric solution to minimize saturation effects in the bright fundamental-mode classical Cepheid, $\beta$~Dor with the lightkurve software, and we revealed strong evidence of cycle-to-cycle variations in the star. In several cases, however, fluctuations in the pulsation could not be distinguished from instrumental effects, such as contamination from nearby sources which also varies between sectors. Finally, we discuss how precise light curve shapes will be crucial not only for classification purposes but also to determine physical properties of these stars.

\end{abstract}

\section{Introduction} \label{sec:intro}

High-precision photometry of Cepheid stars has been of great importance for a long time, as these objects are key to many different areas of astrophysics. Due to the tight relation between the pulsation period and intrinsic brightness, known as Leavitt's Law \citep{leavitt1912}, they are among the primary distance indicators in the Universe. They are also essential in testing pulsation and stellar evolution models. Cepheids encompass a variety of stars: classical Cepheids are young (10$-$100 Myr), intermediate-mass (4$-$9 M$_{\odot}$) supergiants, whereas Type~II Cepheids are old ($>$10
~Gyr), low-mass (0.5$-$0.6 M$_{\odot}$) stars, except for peculiar W Vir stars, which are relatively young
(order of 200 Myr) and probably slightly more massive (0.6$-$0.7 M$_{\odot}$)
than other Type II Cepheids (see \citealt{pilecki2018b}). A third subgroup, anomalous Cepheids, fills the gap at the near-solar mass range. What connects these diverse stars is that they all are crossing the classical instability strip in the Hertzsprung--Russell diagram, pulsate in a single mode or occasionally a few modes of high amplitudes, and follow their respective period-luminosity (P-L) relations. 

Many dedicated photometric measurements targeted the brightest Cepheids from the ground, and furthermore, several large sky surveys searched for Cepheids, or included these stars in their variability catalogs. The Optical Gravitational Lensing Experiment (OGLE, \citealt{ogle-1992}) survey, in particular, observed large numbers of Cepheids both in the Milky Way and the Magellanic Clouds. The decade-long monitoring led to many discoveries, including low-amplitude additional modes and modulation (e.g.~\citealt{Moskalik-2006,smolec-2018,suveges-2018}). Multi-epoch space photometry is also available for a large number of Cepheids with the \textit{Gaia} Data Release 2 (DR2, \citealt{2019Rimoldini}). The \textit{Hubble} Space Telescope observed numerous targets in the Milky Way and other galaxies with a variety of techniques \citep[see, e.g.,][]{hubble-cepheids-2001,rspup-hst-2014,hubble-riess-2018}.

\begin{table*}
\renewcommand{\thetable}{\arabic{table}}
\centering
\caption{Targets analysed in this paper} \label{tab:targets}
\begin{tabular}{lllllll}
\tablewidth{0pt}
\hline
\hline
Type & Name & TIC number & RA & Dec &  $G_\mathrm{RP}$ (mag) & Sector \\
\hline
DCEP-F	&	*$\beta$ Dor	&	149346418	&	83.40630968	&	$-$62.48977125	& 3.26 (T$_\mathrm{mag}$)	&	1,2,3,4,5	\\
DCEP-F	&	RU Dor (OGLE LMC-CEP-4211)	&	277316761	&	84.36790871	&	$-$66.62864148	&	14.045	&	1,2,3,4,5	\\
DCEP-F	&	OGLE LMC-CEP-046	&	294670967	&	70.83688681	&	$-$69.22994857	&	14.090		&	1,2,3,4,5	\\
DCEP-F	&	OGLE LMC-CEP-227	&	30033323	&	73.06543263	&	$-$70.24201219	&	14.626	&	1,2,3,4,5	\\
DCEP-F	&	SX Tuc (OGLE SMC-CEP-3657)	&	182729973	&	16.16673215	&	$-$72.31967806	&	15.540		&	1,2	\\
DCEP-F	&	SW Tuc	(OGLE SMC-CEP-3504)&	182517661	&	15.8665814	&	$-$72.77080623	&	15.548		&	1	\\
DCEP-F	&	AT Tuc	 (OGLE SMC-CEP-0019)&	267184490	&	5.7917735	&	$-$73.52396492	&	15.782		&	1	\\
DCEP-F	&	TT Tuc	(OGLE SMC-CEP-3753)&	52056743	&	16.40772843	&	$-$71.84332273	&	15.953	&	2	\\
DCEP-1O	&	OGLE LMC-CEP-3377	&	33718043	&	61.12042658	&	$-$75.07969818	&	14.698		&	1,2,3,5 	\\
DCEP-1O	&	OGLE SMC-CEP-4955	&	238891904	&	40.62045729	&	$-$74.72155827	&	15.742		&	1,2	\\
DCEP-1O	&	OGLE SMC-CEP-4952	&	50313785	&	31.03923358	&	$-$77.07733787	&	16.971		&	1	\\
DCEP-F/1O	&	OGLE LMC-CEP-4419	&	389744625	&	87.05700922	&	$-$67.55065346	&	15.100		&	2,3,5	\\
DCEP-1O/2O	&	OGLE SMC-CEP-4951	&	50310856	&	30.64132661	&	$-$75.51333933	&	16.803		&	1,2	\\
\hline
ACEP-F	&	*UY Eri	&	9846652	&	48.41317119	&	$-$10.44236242	&	10.759		&	4	\\
ACEP-F	&	OGLE GAL-ACEP-006	&	33877343	&	65.96438605	&	$-$76.91187523	&	12.432		&	1,2, 5	\\
ACEP-F	&	SS Gru	&	129710422	&	322.0262177	&	$-$37.15991246	&	12.557		&	1	\\
ACEP-F	&	DF Hyi	&	24695633	&	25.20499106	&	$-$67.49499412	&	13.888		&	2	\\
ACEP-F	&	WX Tuc	&	234525389	&	15.7575632	&	$-$61.7678435	&	14.200		&	1,2	\\
ACEP-F	&	VV Gru	&	144046456	&	336.863323	&	$-$49.64310665	&	14.700		&	1	\\
ACEP-F	&	AV Gru	&	2027094382	&	334.1492302	&	$-$48.58412015	&	16.844		&	1	\\
ACEP-1O	&	*XZ Cet	&	423761480	&	30.06944318	&	$-$16.34615519	&	8.947		&	3	\\
ACEP-1O	&	AK PsA	(ASAS J215931-3623.5)&	197693113	&	329.8786336	&	$-$36.39168251	&	12.624		&	1	\\
\hline
BLH	&	AA Gru	&	121469834	&	342.4822219	&	$-$46.35802095	&	12.972		&	1	\\
pWVir	&	OGLE LMC-T2CEP-023	&	30473478	&	75.05428248	&	$-$67.71216698	&	15.677		&	1,3,4,5	\\
pWVir	&	OGLE LMC-T2CEP-280	&	141824491	&	93.22641561	&	$-$73.73221672	&	15.853		&	1,2,3,4,5	\\

\hline
Non-Cepheid	&	RV Men	&	141870888	&	94.29857588	&	$-$73.48165286	&	10.784		&	1,2,3,4,5	\\
\hline
\end{tabular}
\tablecomments{Abbreviations for variable types: DCEP-F (fundamental-mode classical Cepheid), DCEP-1O (first-overtone classical Cepheid), DCEP-F/1O (double-mode classical Cepheid pulsating in fundamental and first overtone modes), DCEP-1O/2O (double-mode classical Cepheid pulsating in first and second overtone modes), ACEP-F (fundamental-mode anomalous Cepheid), ACEP-1O (first-overtone anomalous Cepheid), BLH (BL Her type), pWVir (peculiar W Vir type).  
Literature variable types were revised for OGLE SMC-CEP-4952 and RV Men. 2-minute cadence targets are marked with symbol *. No $G_\mathrm{RP}$ value is available yet for $\beta$~Dor,  therefore we gave the \textit{TESS} brightness (T$_\mathrm{mag}$).}
\label{table:targets}
\end{table*}

Continuous space-based light curves gathered over extended periods of time, ranging from weeks to years, led to multiple discoveries concerning the pulsation properties of Cepheids. Observations made with the star tracker camera of the \textit{WIRE} space telescope and with the \textit{SMEI} instrument on the \textit{Coriolis} satellite revealed that the pulsation of Polaris is returning after reaching minimum amplitude more than a decade earlier \citep{bruntt-2008}. The \textit{Kepler} space telescope detected both very small modulations and low-level, short-timescale instabilities that appear in the form of cycle-to-cycle change and period jitter in V1154~Cyg, the only classical Cepheid in the original field \citep{derekas-2012,derekas-2017}. In contrast, other classical Cepheids observed by the \textit{CoRoT} space telescope remained stable with clockwork precision \citep{poretti-2015}. Data from the \textit{MOST} space telescope indicated that the pulsation of the first-overtone star SZ~Tau is less stable than that of the fundamental mode star RT~Aur \citep{evans-2015}. Period doubling and weak signs of additional non-radial modes were also found in V473~Lyr and U~TrA, respectively, with \textit{MOST} \citep{molnar-2017}. So far the \textit{BRITE} satellites observed the largest number of Cepheids, and detected phenomena ranging from low-amplitude modulation cycles to extra modes and jitter in them \citep{brite-2017,brite-2018}. 

Some Type~II Cepheids (T2CEP) are known to exhibit non-linear dynamical phenomena \citep{2016Smolec, smolec-2018}. These are believed to be the results of destabilization of the fundamental mode by non-linear coupling to another pulsation mode. The secondary mode remains hidden since it is not excited with an amplitude large enough to be detected, and/or is locked into resonance with the fundamental mode. Non-linear phenomena may appear as period doubling or chaos, the latter only manifesting in long period Type~II Cepheids, also known as RV~Tau variables. Chaotic behavior was confirmed both from the ground \citep{buchler-1996,kollath-1998} and via \textit{Kepler} data \citep{dfcyg1,dfcyg2}. Period doubling was also detected both with ground- and space-based observations in each subclass of Type II Cepheids \citep{smolec2012,2017Plachy, smolec-2018, jurkovic2020}. We note that light curves of Type II Cepheids can also be affected by extrinsic variations, usually connected to binarity. This is displayed most prominently in the RV~Tau stars with long secondary variations, caused by obscuration from a circumbinary disk \citep[see, e.g.,][]{dfcyg1,kiss-bodi-2017,vega-2017}. Slow variations in the light curve of the W~Vir-type star IX~Cas may also indicate the presence of near-side heating and eclipses happening in the system \citep{turner-2009}.

Non-linear phenomena are more rare among classical Cepheids. So far only one star was found to show period doubling: the second-overtone classical Cepheid, V473~Lyr. This star is known for its strong amplitude and phase modulations \citep{molnar-2014}. Modulation is common among RR~Lyrae stars: about half of them show the phenomenon called the Blazhko effect \citep[see, e.g.,][]{jucsik-2009}, but less so among Cepheids. Nevertheless, modulation sometimes appears with a moderate amplitude in classical Cepheids, and it was also identified in stars spanning all subtypes of Type~II Cepheids \citep{2017-smolec, smolec-2018}. Modulations and additional modes are also found in time series radial velocity (RV) measurements  \citep{anderson2014,anderson2019}. 
\citet{stothers-2009} proposed a possible mechanism to explain the period and amplitude changes in the pulsation of short-period Cepheids. His suggestion involves magnetoconvective cycles that periodically weaken and invigorate convection which in turn eventually affects both the pulsation amplitude and period. However, one would need complex magnetohydrodynamic modeling to confirm this hypothesis. Hydrodynamic models also suggest that this mechanism cannot produce fast modulation \citep{smolec-2011,molnar-2012}. More recently, mode resonances arose as viable explanation for the Blazhko effect in RR~Lyrae stars: such a mechanism was reproduced in BL Her models and might be able to arise in classical Cepheids as well \citep{buchler2011,smolec-2012}.

The 2018 launch of the NASA Transiting Exoplanet Survey Satellite (\textit{TESS}) has brought forward a new opportunity to expand Cepheid science. The extensive observed area and magnitude range of the \textit{TESS} mission facilitates observation of all Cepheids in the Milky Way as well as the brighter ones in the Magellanic Clouds. 

In this paper, we present the first results on Cepheid stars observed with \textit{TESS}, covering the first 5 out of 26 sectors of its primary two-year near all-sky mission. We show that the pulsation properties presented above are detectable with \textit{TESS} as expected, albeit with some limitations imposed by the short time interval covered by observations and strong blending. In Section 2 we introduce the photometric solution we used and explain our selection of targets. In Section 3 we present our analysis along with the results. We also discuss the relevance of our findings and our experiences with \textit{TESS} data including the instrumental issues that we faced. Section 4 includes a short summary and conclusions.

\begin{figure*}
\begin{center}
\noindent
\resizebox{57mm}{!}{\includegraphics{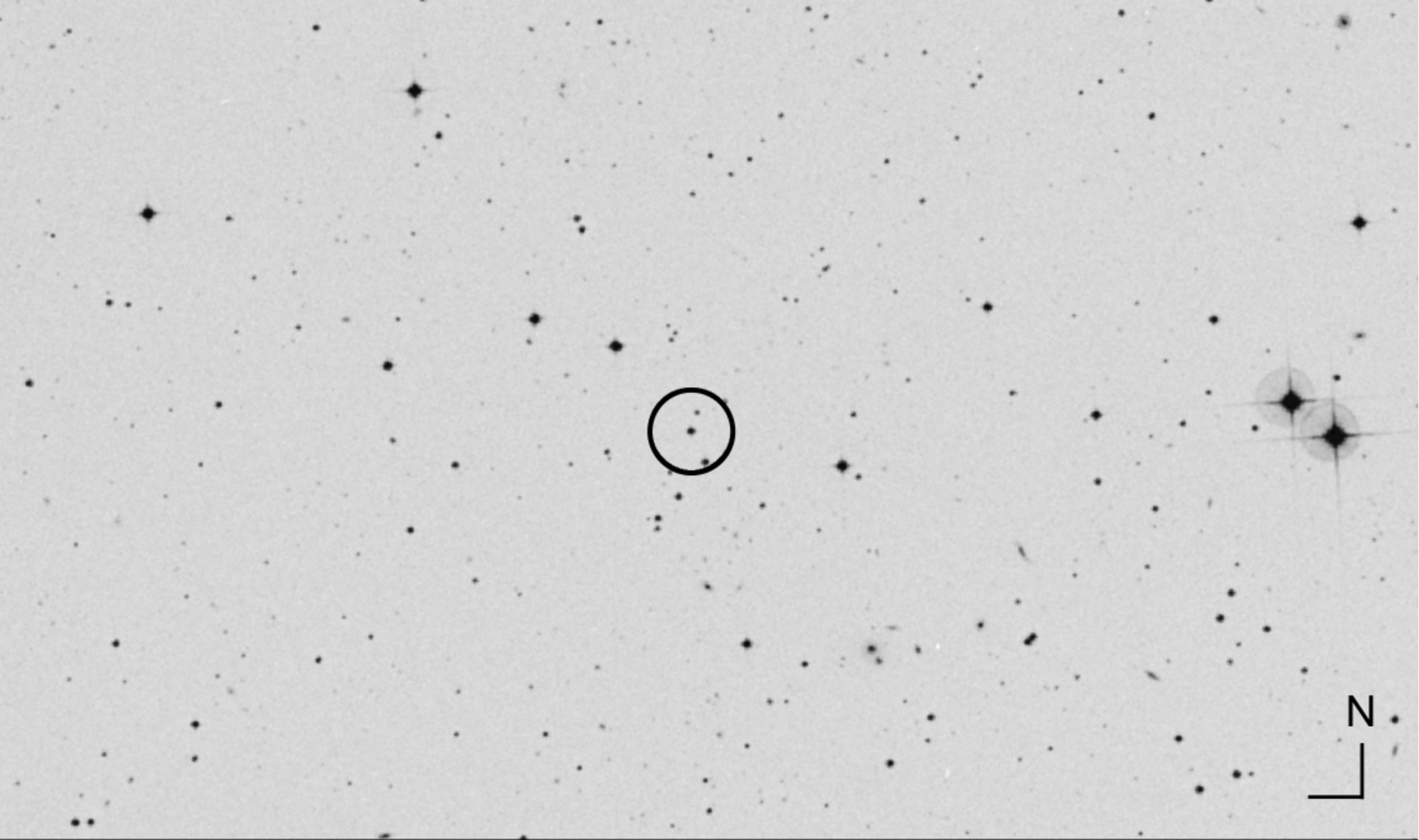}}\hspace*{2mm}%
\resizebox{57mm}{!}{\includegraphics{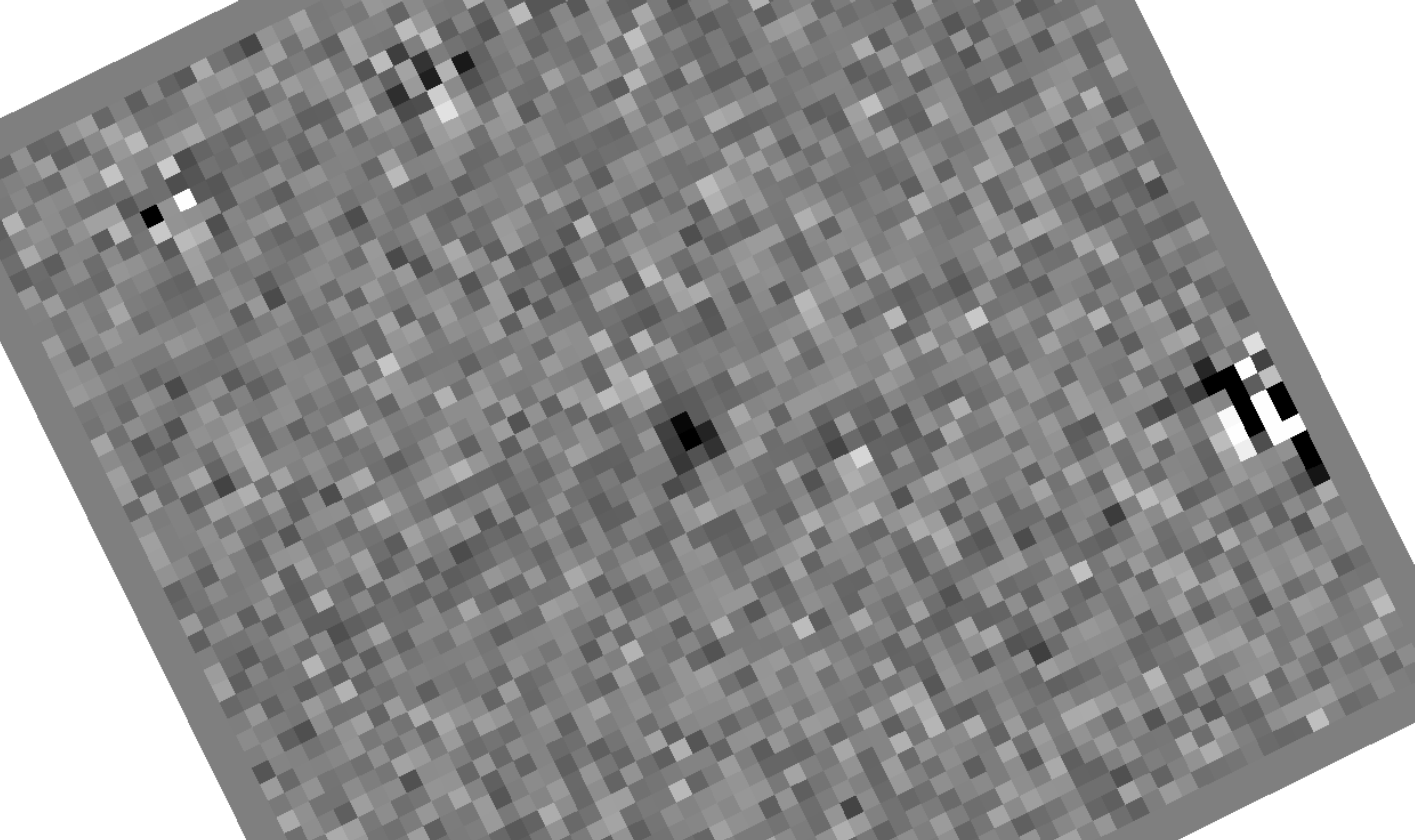}}
\resizebox{57mm}{!}{\includegraphics{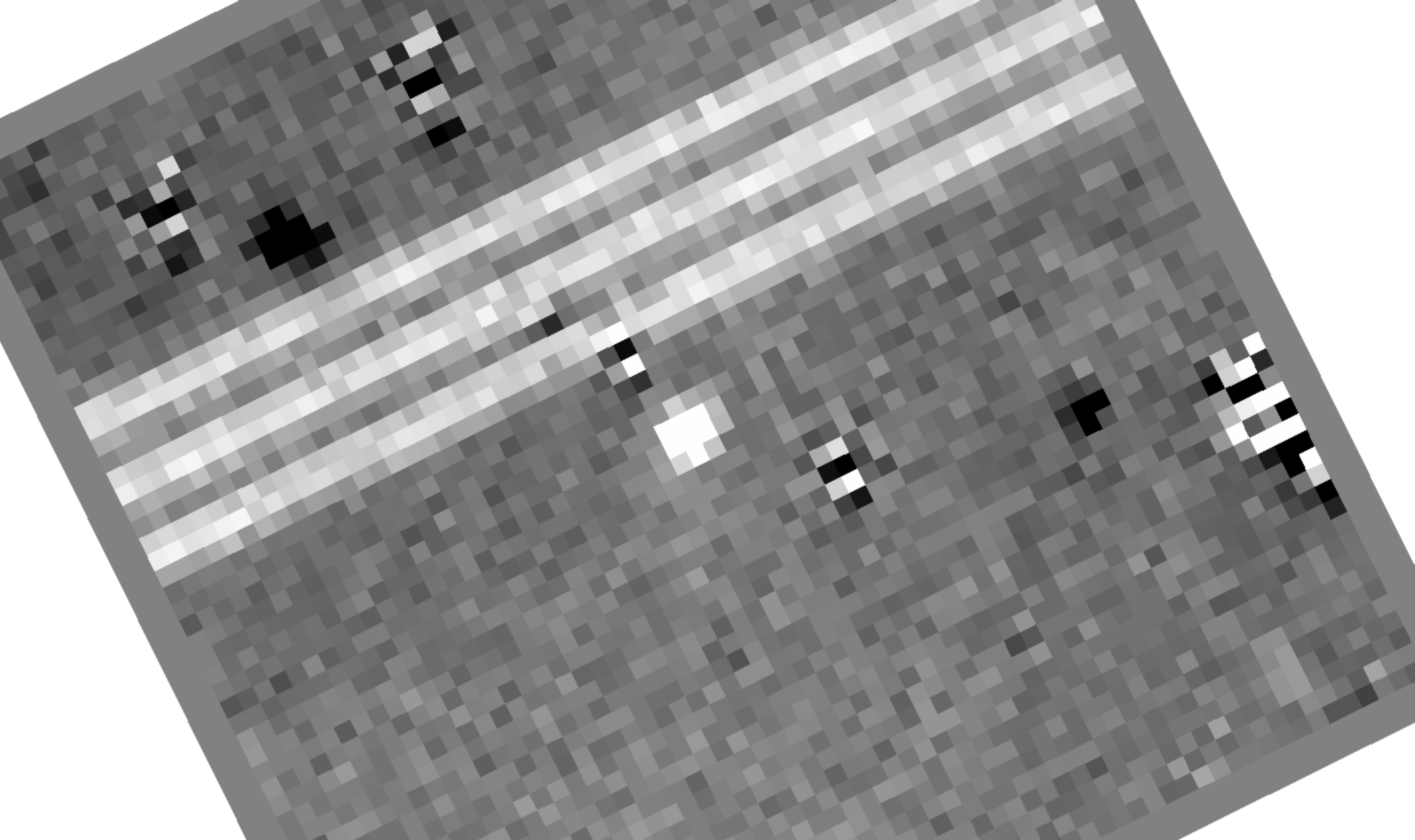}}\hspace*{2mm}%
\vspace*{2mm}

\noindent
\resizebox{57mm}{!}{\includegraphics{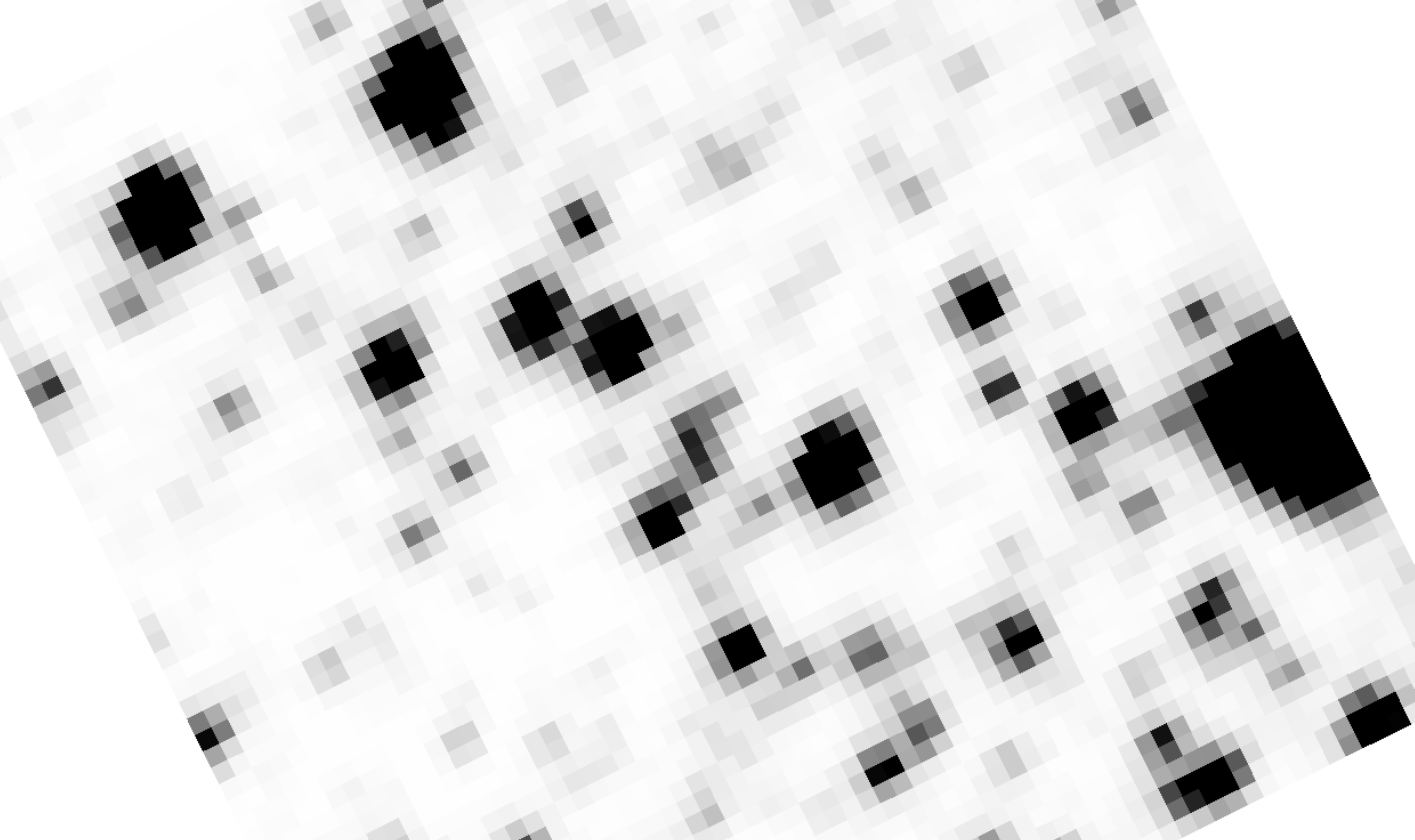}}\hspace*{2mm}%
\resizebox{57mm}{!}{\includegraphics{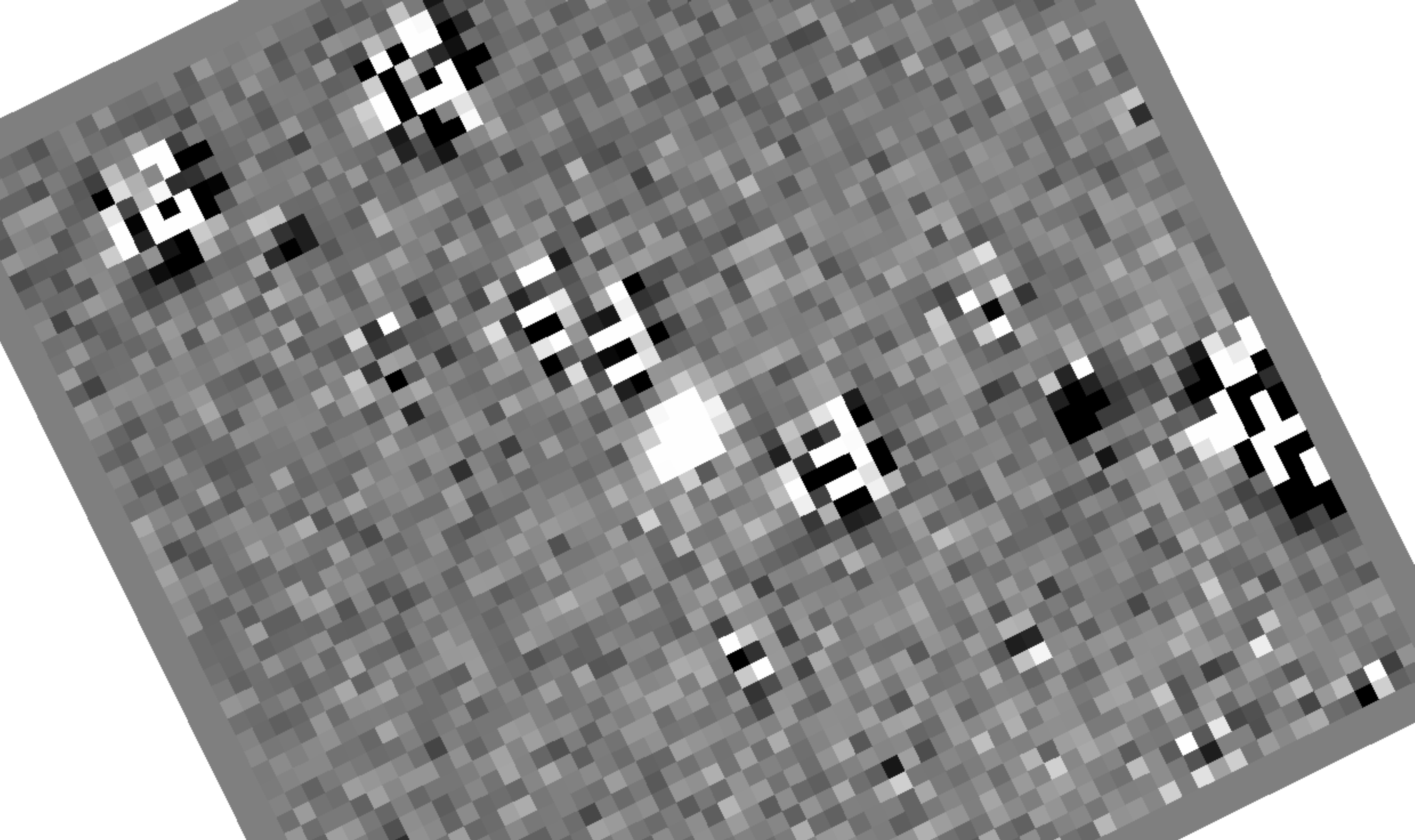}}\hspace*{2mm}%
\resizebox{57mm}{!}{\includegraphics{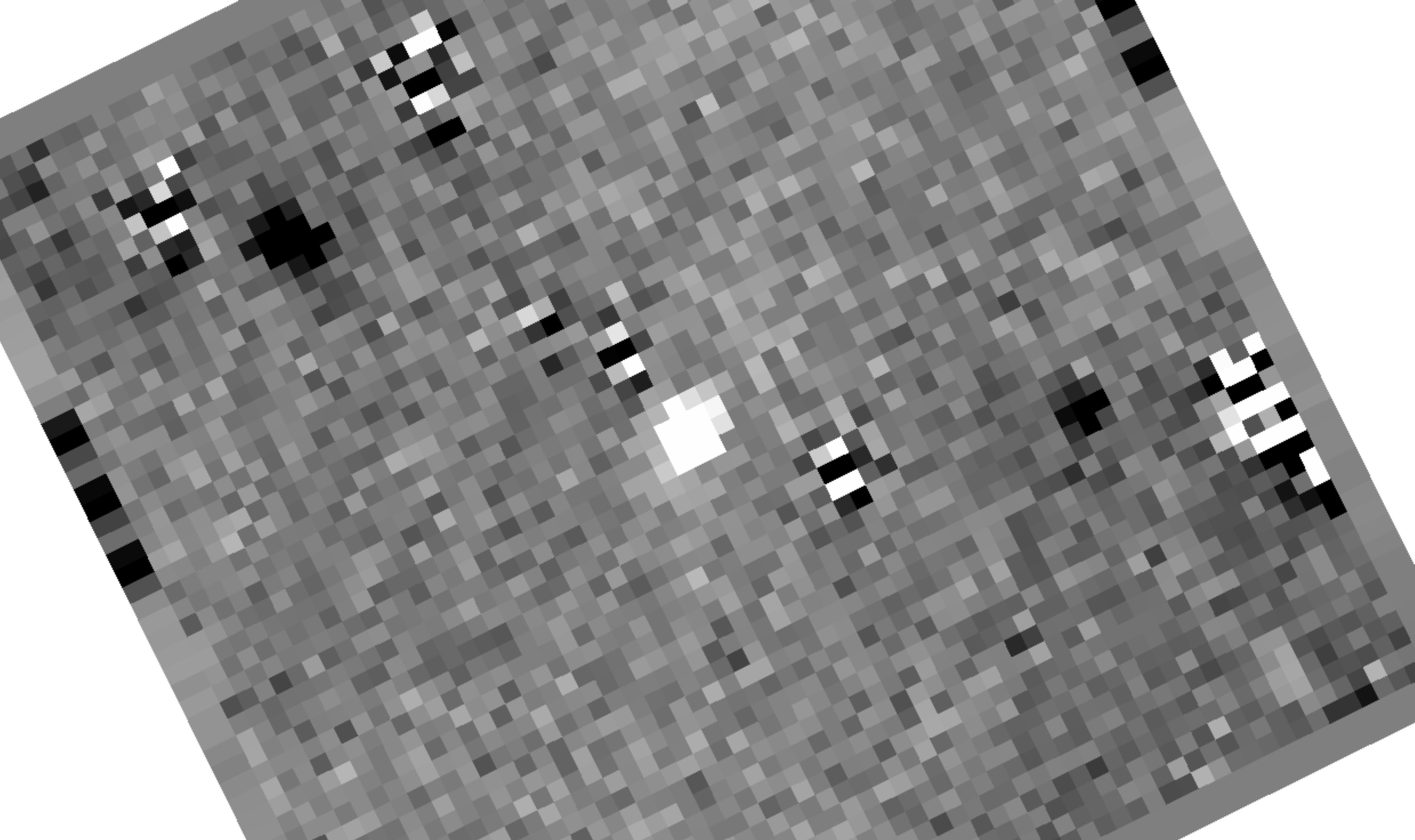}}
\end{center}
\caption{Images of the target star WX Tuc throughout the various steps of the processing. 
\emph{Upper left:}
DSS image of the star (black circle) and its surroundings.
\emph{Lower left:} 
The corresponding reference \textit{TESS} image, the median of
$11$ individual trimmed full-frame images. 
\emph{Upper middle:}
A differential image after applying the best-fit image convolution
transformation. The differential
velocity aberration was minimal at this image. 
\emph{Lower middle:}
A differential image after applying the best-fit image convolution
transformation. Here the differential
velocity aberration was rather significant, therefore the residual structures
are quite prominent around the brighter (8-9 mag) stars. 
\emph{Upper right:}
A differential image with prominent stripes induced 
by stray Earth light reflecting from CCD electronics.
\emph{Lower right:}
The same image as shown in the upper right panel but after removing the 
stripes. 
}
\label{fig:wxtuctessdiff}
\end{figure*}

\section{Targets and data} \label{sec:data}
\textit{TESS} observations consist of $\sim$27 day long measurements with constant pointing to a 24$^\circ\times$  96$^\circ$ field of view, called a Sector \citep{ricker2015}. Sectors are rotated around the Ecliptic Pole, creating overlapping regions around the pole. Sectors are made up of two consecutive orbits of the spacecraft around the Earth, each with a period of half a lunar sidereal month. Measurements are continuous, except for a few hours at perigee when data are being downloaded. A selected set of targets is observed in 2-minute cadence mode with pre-defined pixel masks, while full-frame images (FFIs) are stored with 30-minute exposure time. FFIs consist of 16 individual 2k$\times$2k frames from the 4 CCDs of each of the 4 \textit{TESS} cameras. Photometric data can be obtained for all objects in FFIs to a brightness limit of $\sim$16-17 mag in the \textit{TESS} filter, which has a spectral response function of a broad passband (600-1000 nm) centered on the Cousins $I$-band \citep{ricker2015}. The light curves of stars located in dense stellar fields are affected by strong contamination from the nearby stars due to the low angular resolution that is determined by the 21$\arcsec$ pixel size. FFIs and the 2-minute cadence light curve products created by the \textit{TESS} Science Processing Operations Center (SPOC, \citealt{jenkinsSPOC2016}) are publicly available at the Mikulski Archive for Space Telescopes (MAST\footnote{\url{https://archive.stsci.edu/}}) and continuously updated with the data from consecutive sectors.  

\textit{TESS} was designed to monitor over 85\% of the sky in the primary mission, covering the majority of Galactic and nearby extragalactic Cepheids, but the observational limitations mentioned above restrict us to a subset of optimal Cepheid targets. The success of our analysis depends on both the photometric precision and the number of observed pulsation cycles. Therefore the brightness, the pulsation period and the celestial position play equally important roles in selecting the best targets \citep{plachy2020}. The aim of this study is to test the data quality and 
to assess the possibilities and the limitations the \textit{TESS} observations may offer for various subtypes of Cepheids. We chose our sample targets from the early sectors of \textit{TESS}. Three of them were observed in 2-minute cadence mode: the bright classical Cepheid $\beta$~Dor, which is located in the southern continuous viewing zone (CVZ) of \textit{TESS}, and two anomalous Cepheids, XZ~Cet in Sector 3 and UY~Eri in Sector 4. The rest of our stars are FFI targets. The early sectors avoided the Galactic disk where classical Cepheids are generally located, thus the majority of our Galactic sample consist of Type~II and anomalous Cepheids that are more evenly distributed in the sky and can be found in the halo too. Along with the Galactic targets we selected Cepheids from the Magellanic Clouds and Bridge, as well. This area is monitored and well-studied by the OGLE Survey that provided us with classical Cepheid targets of different pulsation modes. Moreover, the Large Magellanic Cloud (LMC) lies in the CVZ of \textit{TESS}, thus the targets here were observed over multiple sectors.

The selected targets are listed in Table \ref{table:targets}. Some stars were observed in more than one sector depending on the overlap of the fields of view. Since no light curves are provided for the FFI targets, we decided to generate our own photometry. Below we describe how we created the light curves before proceeding with the analysis. 

\subsection{Data reduction with FITSH}
\label{sec:reduction}
To produce light curves from the FFIs we used the FITSH package developed by \citet{pal2012}. This tool provided us differential photometric solutions for the target stars. The data reduction process as implemented in the FITSH pipeline has been split into two parts, namely the global full-frame analysis and the target object photometry. 
As the first step, the calibrated FFIs are analyzed in terms of derivation of the plate solution. We perform the astrometric cross-matching with the \textit{Gaia} DR2 source catalog \citep{gaia2016,gaia2018}. In this step, we also derive the 
flux zero-point with respect to the \textit{Gaia} $G_\mathrm{RP}$ magnitudes due to the similarities between the \textit{TESS} and the \textit{Gaia} photometric systems. Finalizing the plate solutions, corrections were applied to account for all other optical aberrations as well as to treat the differential velocity aberration.

In the second step, we trim small, 64$\times$64 pixel frames centered on our target stars and execute the differential image analysis only on those sub-frames. Any type of differential image analysis requires a so-called reference frame, from which the deviations are expected to be minimal and signal-to-noise (S/N) ratio to be maximal. To create this reference frame, we select 9 or 11 individual and subsequent sub-frames for each sector around the mid-times of the observation series when differential velocity aberration, and thus the apparent shifts in the stellar images are minimal. We calculated the median of these frames to maximize the S/N. 

We needed to obtain a set of reference fluxes for the targets of interest in parallel with the derivation of the image convolution coefficients. The image convolution is able to correct for many of the instrumentation and/or intrinsic differences between the target frames and the reference frame, including the slight drift caused by the differential velocity aberration, the spacecraft jitter, background and stray light variations. It also helps to eliminate or significantly decrease the effects on the frames acquired during momentum wheel dumps. Since reference fluxes are difficult to accurately obtain even at moderately confused stellar fields for \textit{TESS} due to the large pixel size, we simply used the \textit{Gaia} DR2 $G_\mathrm{RP}$ magnitudes for this purpose. However, the final fluxes (i.e., the sums of the reference fluxes and the respective residual fluxes on the convolved and subtracted images) needed to be adjusted because the phase of the light variations could be significantly different in the reference frames assembled for the subsequent \textit{TESS} sectors. The aperture size was set to be 2.5 pixels (52.5 $^{\prime\prime}$), which worked well for most of our targets. We tested smaller aperture sizes (1.5 px and 2.0 px) and found that they have no effect on the frequency content but can reduce the measured pulsation amplitude by up to 30\%. Therefore we used 2.5 px apertures whenever possible, except for very crowded fields where we used smaller radii to minimize contamination from nearby variable stars. In Fig.~\ref{fig:wxtuctessdiff}, we display example frames for WX~Tuc, including a Digitized Sky Survey image, the reference image used for the differential photometry process as well as some typical differential frames showing
up during the process.

The light curve outputs of the FITSH pipeline contained a small number of outliers, which are mostly due to momentum dumps of the reaction wheels at every 2.5 days when \textit{TESS} is not in fine pointing for a few minutes. Low-quality segments can also be identified in some light curves that are caused by other issues, such as stray light or glints. To eliminate these data points we used a 3$\sigma$-clipping cleaning method: we deleted points that differed more than three times the standard deviation of the residual compared to a fitted signal composed of the five highest-amplitude frequencies.  Finally, we applied barycentric correction to the times stamps of each target. The light curves are provided in the Appendix and are also available online for further analysis\footnote{\url{https://konkoly.hu/KIK/data\_en.html}}.

\begin{figure}
\begin{center}
\resizebox{40mm}{!}{\includegraphics{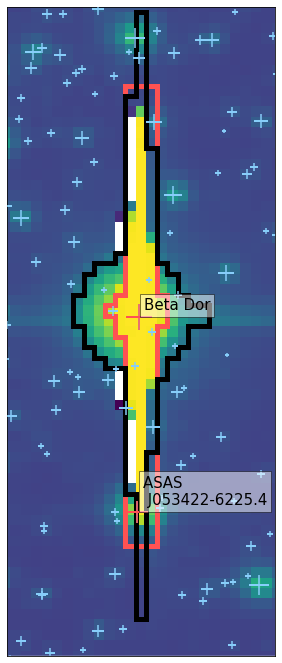}}\hspace*{0mm}%
\resizebox{40mm}{!}{\includegraphics{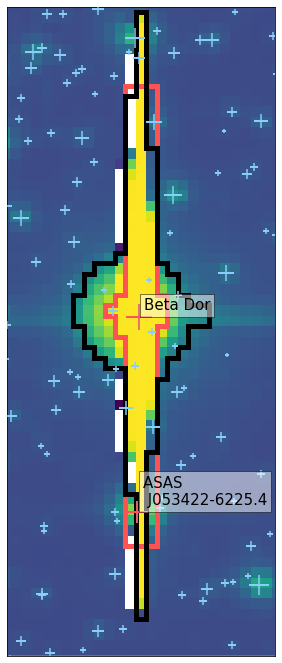}}\hspace*{0mm}%
\end{center}
\caption{The SPOC pixel mask (red) and our custom pixel mask (black) for $\beta$~Dor in Sector 1. Note the difference in the length of the bleed columns in minimum (left) and maximum light (right): the latter exceeds the SPOC mask.}
\label{fig:betadormask}
\end{figure}
\begin{figure*}
\begin{center}
\includegraphics[width=1.0\textwidth]{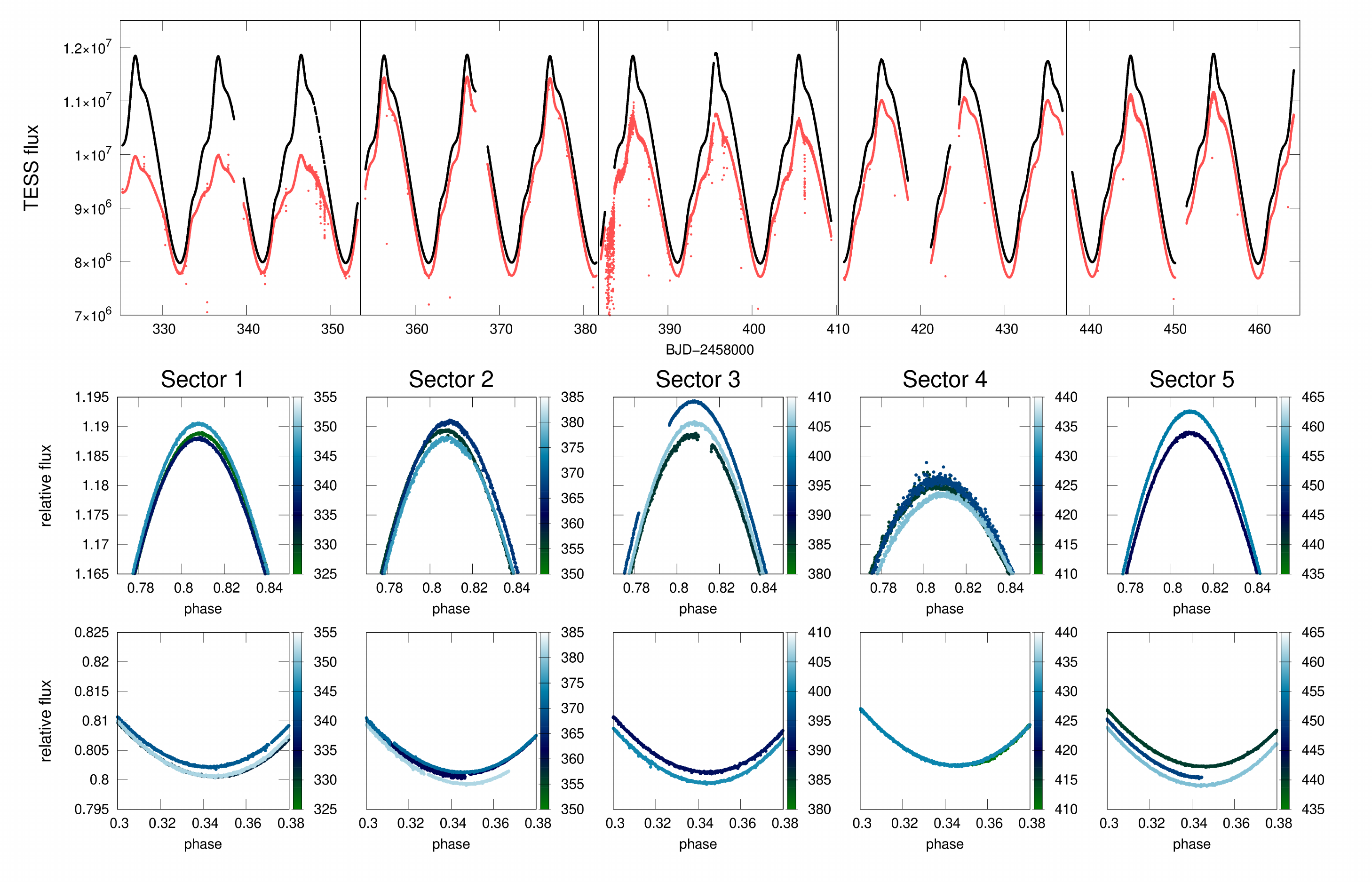}
\end{center}
\caption{2-minute cadence photometry of $\beta$~Dor from the SPOC pixel mask (red) and our custom pixel mask (black). Lower panels show the differences in the maxima and minima of the custom light curves when folded with the pulsation period, for each sector. Colours indicate the progress of time (BJD$-$2458000) in days.}
\label{fig:betadorlc}
\end{figure*}
\subsection{2-minute cadence data} \label{sec:short_cadence}

Three of our stars were included in the 2-minute cadence target list assembled by the \textit{TESS} Asteroseismic Science Consortium: $\beta$~Dor, XZ~Cet, and UY~Eri. In the cases of XZ~Cet and UY~Eri we downloaded the SPOC-generated light curves from MAST, and used the Simple Aperture Photometry, \texttt{SAP\_FLUX} \citep{twicken:PA2010SPIE,morris2020},  that we converted to magnitude units using 20.54 mag as the zero point. We applied outlier removal to them, and proceeded with the analysis. However, we found that the SPOC light curve for $\beta$~Dor produced very different flux values from one sector to the other. As $\beta$~Dor is bright enough to saturate the \textit{TESS} detectors and it is a large-amplitude variable, the length of the bleed column (i.e. the set of pixels that store the excess flux from the star) changes considerably over the pulsation cycle. Closer inspection revealed that the photometric aperture set by the pipeline was not always capturing the bleed columns entirely. This problem cannot be solved with scaling the light curves of different sectors to the same amplitude, because flux loss occurs only in the brightest pulsation phases (i.e. around the maxima), and therefore the light curve gets periodically distorted. We note that the same variable bleed column length problem affected the observations of the star RR~Lyr during the early stages of the \textit{Kepler} mission \citep{kolenberg2011}.

In order to capture the flux from the star properly, we generated custom apertures that were optimized to the bleed columns at maximum light with the \texttt{lightkurve} tool \citep{lightkurve1,lightkurve}. Fortunately, the target pixel files (TPFs) were large enough to contain the bleed column entirely in 4 out of 5 sectors, albeit barely. The comparison of the SPOC and the custom masks of Sector 1 is presented in Fig.~\ref{fig:betadormask}. Even with the custom apertures, small differences between the successive sectors still remained. We were able to minimize these by scaling and shifting the sectors to the same median flux and amplitude, except for Sector 4 where we detected flux loss caused by the bleed column extending beyond the TPF edge at maximum light. The sector-stitched SPOC and custom-aperture light curves can be seen in Fig.~\ref{fig:betadorlc}, as well as the phased light curves of each sector that we discuss in Section \ref{sec:betador}. Our custom light curve of $\beta$~Dor is also available in the Appendix and online, together with the FITSH reduced ones.

\section{Analysis and results} \label{sec:analysis}

We performed standard Fourier analysis on the targets. We determined the main pulsation frequency and its harmonics (or frequencies and linear combinations for double-mode stars) with the detection limit of S/N$>$4. We searched for possible peaks of low amplitude additional modes or modulation. We compared the results calculated by different tools: \texttt{Period04}, \texttt{MuFrAn}, \texttt{LCfit} \citep{period04,mufran,lcfit} as well as custom codes. 
The main pulsation frequencies obtained from each approach agree to a level of approximately $10^{-4}-10^{-5}\,\mathrm{d}^{-1}$ for the main frequencies. We also calculated the Fourier parameters $R_{21}$, $R_{31}$, $\phi_{21}$ and $\phi_{31}$, the relative amplitude and phase values as defined by \citet{simon-lee-1981}: $R_{i1}$=$A_i/A_1$ and $\phi_{i1}$=$\phi_i - i\phi_1$, where $i=2, 3$.  These parameters have great importance in the classification of sub-types that we have revised and discuss in Section \ref{sec:foupar}. They can also be the key in the estimation of theoretical masses, as shown in Section \ref{sec:masses}. Table~\ref{table:fou} contains the results of our analysis, the main period and frequency, the harmonic series, combinations and additional peaks. The uncertainties are calculated with the least squares fit method using \texttt{Period04}. 

\begin{table*}[ht]
 
\renewcommand{\thetable}{\arabic{table}}
\centering
\caption{Summary table of the results.} \label{tab:results}
\begin{tabular}{llllll}
\tablewidth{0pt}
\hline 
\hline
\textbf{Name} & \textbf{Period} (d) &  \textbf{Freq.} (d$^{-1}$) & \textbf{Harm./Comb.} &$R_{21}$, $\phi_{21}$, $R_{31}$, $\phi_{31}$ & \textbf{Comments} \\
\hline
\multicolumn6l{\textbf{Fundamental-mode Classical Cepheids}}\\
\hline
$\beta$ Dor & 9.84318(2) &  0.1015932(2)&...10$f$,12$f$ 	& 0.069, 5.425, 0.104, 3.374 & 	\\
RU Dor 	& 	8.3259(4) &	0.120107(6)	&	...9$f$,11$f$ & 0.307,	3.770, 	0.228,	0.626 & contamination	\\
LMC-CEP-046 & 8.8437(2) &	0.113075(3)	&	...9$f$ & 0.235,	3.499,	0.194,	0.625 &  contamination	\\
LMC-CEP-227 & 3.7972(4)	 &	 0.26335(3)	&...9$f$	& 0.490,	2.858,	0.280,	5.775 & contamination	\\
SX Tuc & 3.3530(6)	 & 0.29824(5)	& ...8$f$	& 0.505, 2.723, 0.317, 5.615 & strong blend 	\\
SW Tuc 	& 3.5914(14) & 0.27844(11)	& ...9$f$ & 0.482,	2.721,	0.296,	5.509 & contamination\\
AT Tuc 	& 	3.1155(14) &	0.32097(14)&	...9$f$ & 0.506,	2.628,	0.292,	5.311 & contamination   \\
TT Tuc 	&  2.3463(8) &	0.42616(15)	& ...9$f$ & 0.543, 2.596,	0.386,	5.430 & blend\\
\hline
\multicolumn6l{\textbf{First-overtone Classical Cepheids}}\\
\hline
LMC-CEP-3377& 3.21455(8)	 &	0.311085(8)	& ...4$f$,  $f_x$=0.5054(4) 	& 0.179, 2.151, 0.074, 3.606
 & contamination	\\
 SMC-CEP-4955& 2.0313(3)	 &	0.49228(7)	& ...3$f$	& 0.161, 3.134, 0.030, 0.671
 & contamination	\\
SMC-CEP-4952& 	1.6420(13) &	0.6090(5)	& ...2$f$	& 0.242, 2.670
 & 	noisy\\

\hline
\multicolumn6l{\textbf{Double-mode Classical Cepheids}}\\
\hline

LMC-CEP-4419& 	$p_\mathrm{F}$=3.4790(5) &	$f_\mathrm{F}$=0.28744(4)& $f_\mathrm{F}$, $2f_\mathrm{F}$, $f_\mathrm{1O}$,   & 0.170, 2.822 &	blend\\
& $p_\mathrm{1O}$=2.4942(4)	 &	$f_\mathrm{1O}$=0.40093(7)&    $f_\mathrm{F}$+$f_\mathrm{1O}$, $2f_\mathrm{F}$+$f_\mathrm{1O}$ & 	&\\ 
SMC-CEP-4951&  $p_\mathrm{1O}$=0.71708(5) & $f_\mathrm{1O}$=1.39453(10) & $f_\mathrm{1O}$, $2f_\mathrm{1O}$, $3f_\mathrm{1O}$, 	& 0.263, 2.281, 0.111, 5.162
  &  noisy	\\
& $p_\mathrm{2O}$=0.5777(2)  &  $f_\mathrm{2O}$=1.7308(6)	& $f_\mathrm{2O}$, $f_\mathrm{1O}$+$f_\mathrm{2O}$	&  & 	\\
\hline
\multicolumn6l{\textbf{Fundamental-mode anomalous Cepheids}}\\
\hline
UY Eri & 2.21357(2) & 0.451758(4)	& ...24$f$	& 0.415, 3.281, 0.159, 0.174 & slight trend	\\
GAL-ACEP-006& 1.883636(7)	 &0.530888(2)	&...19$f$ & 0.410, 3.107, 0.192, 6.261& incoherent peaks	\\
SS Gru & 0.959800(3) & 1.041884(4)	& ...22$f$ (Nyquist)	& 0.464, 2.688, 0.317, 5.283& 	\\
DF Hyi &1.122614(7)  &0.890778(5)	& ...17$f$ 	& 0.489, 2.574, 0.346, 5.179& 	\\
WX Tuc & 0.837492(7) &1.194044(10)	& ...18$f$	& 0.396, 2.537, 0.277, 4.867 & trend	\\
VV Gru &0.826377(10)  &1.21010(2)	& ...15$f$	& 0.353, 2.509, 0.265, 4.789& 	\\
AV Gru & 1.00581(10) & 0.99422(10)& ...8$f$	& 0.479, 2.575, 0.270, 5.172&  noisy	\\
\hline
\multicolumn6l{\textbf{First-overtone anomalous Cepheids}}\\
\hline
XZ Cet & 0.822974(5) & 1.215105(8)	&...13$f$, see Table \ref{table:xz}	& 0.172, 3.621, 0.087, 0.781 & 	\\
AK PsA & 0.82210(10)	 & 1.21623(2)	& ...9$f$, &0.151, 2.884, 0.047, 6.050 & incoherent peaks	\\
\hline
\multicolumn6l{\textbf{Type~II Cepheids}}\\
\hline
AA Gru & 2.45230(10)  &0.40778(2)	&  ...16$f$	& 0.183, 3.712, 0.120, 0.269& 	\\
LMC-T2CEP-023 & 5.2312(8) & 0.19116(3)& ...$8f$, comb. peaks	& 0.297, 3.600,	0.142, 	0.219
 & binary, blends	\\
LMC-T2CEP-280 &5.4162(3)  &0.184627(10) & 	...13$f$&0.376, 3.644, 0.201, 0.913
  & 	\\
  \hline
\multicolumn6l{\textbf{Non-Cepheid}}\\
\hline
RV Men & 0.542250(3) & 1.844169(10)& ...6$f$&0.264, 2.331,	0.027, 1.120  & incoherent peaks	\\
\hline

\end{tabular}
\label{table:fou}
\end{table*}

We also investigated the stability of the pulsation.
Variations in the O$-$C diagrams, a power excess in the residual frequency spectra, and the apparent differences in the phased light curves may all indicate intrinsic instabilities of the pulsation. However, instrumental effects could also be responsible for these features, similarly to excess variation originating from contaminating sources that may also vary between sectors. Distinguishing between small intrinsic variations and instrumental or blend effects is a real challenge and it is impossible in many cases. Nevertheless, we are able to present some cases where the intrinsic origin is the most likely explanation for the observed instabilities. 

Results are presented in the following subsections in detail.

\subsection{$\beta$ Dor}
 \label{sec:betador}

The star $\beta$~Dor is among the brightest classical Cepheids in the sky, therefore, it has been the main subject of several earlier studies \citep[see, e.g.,][]{bell1967,lub1979,turner1980}. 
 It is also an important contributor to the Cepheid-based investigations concerning the Galactic structure and rotation \citep{betador-kraft1963}, its abundance gradient \citep{andrievsky2002}, the P-L relation \citep{benedict2007}, the pulsation velocity projection factor  \citep{parsons1972,nardetto2006,nardetto2007,breitfelder2016}, to mention just a few. The \textit{TESS} data of $\beta$~Dor provide new and precise input for models and future analysis.

Here we focus on only one phenomenon that is clearly visible in the light curve in Fig.~\ref{fig:betadorlc}: the cycle-to-cycle variation of the amplitudes. This is seen not only between sectors but during a particular sector, too, ruling out incorrect data stitching as a cause. The variation reaches $\sim$5$\times$10$^4$ e$^-$/s flux ($\sim$5 mmag, $\sim$1.3 percent of the peak-to-peak amplitude), which is two orders of magnitude higher than the photometric precision of the data points. Blends cannot be responsible for these differences either: only one neighboring bright star shows high-amplitude variation, ASAS~J053422-6225.4 ($G_{\rm RP} = 7.95$ mag), but on a much longer timescale than the observed phenomenon. Moreover it contaminates only in Sector 1, when it falls in the saturation column (Fig.~\ref{fig:betadormask}), but not in later sectors. Thus we conclude that the cycle-to-cycle variations must be intrinsic, and $\beta$~Dor is another example where precise space-based photometry reveals low-amplitude fluctuations in the pulsation, after SZ~Tau, observed by the \textit{MOST} satellite \citep{evans2015}, and V1154~Cyg, measured with \textit{Kepler} \citep{derekas-2017}. Cycle-to-cycle variation in the RV curves has been detected in the long-period fundamental-mode Cepheids RS~Pup  and $\ell$~Car, on the order of  a few km\,s$^{-1}$ \citep{anderson2014,anderson2016,anderson2016b}. Evidence for cycle-to-cycle differences in the angular diameter variability has also been found in the case of the latter star, which may lead to $\sim$10 mmag photometric fluctuations.    
For $\beta$~Dor, an upper limit of $< 0.57$ km\,s$^{-1}$ was found for cycle-to-cycle variations in the RV curve by \citet{taylor-booth-1998}. 

The investigation of RV and photometric fluctuations in Cepheid pulsations is a new and promising area of research. As we showed in the data of the early sectors of $\beta$~Dor, \textit{TESS} makes it possible to combine accurate photometric observations with spectroscopic measurements of modulation and fluctuations in the pulsation for the brightest Cepheids in the sky and thus has the potential to provide important contributions to solve the question of (in)stability of Cepheid pulsation. 

\subsection{Galactic Anomalous and Type~II Cepheids} \label{sec:targets}

Anomalous Cepheids (ACs), also known as BL~Boo stars form the rarest and least understood type of Cepheids. They are $\sim$0.5-2 mag brighter than RR~Lyrae stars of similar pulsation periods. This property led to their discovery in the Sculptor dwarf galaxy \citep{baade-hubble}. Since then dozens of anomalous Cepheids have been found in the Local Group dwarf galaxies (see \citealt{martinez-vazquez, stetson} and references in \citealt{fiorentino}) and hundreds in the Magellanic Systems \citep{anom-ogle, anom-ogle2}. More than fifty ACs are known in the Galactic bulge \citep{acep-bulge1, udalski18} but only a few in the Galactic globular clusters \citep{zinn-dahn, navarrete}. \citet{mccarthy1997} presented the only detailed high-dispersion spectroscopic study of an anomalous Cepheid (BL~Boo) so far.

Single-star evolution models indicate that ACs are intermediate-age ($1-6$ Gyr), metal-poor (Z$=\sim0.0001$), He-burning stars with masses of 1.3--2M$_{\odot}$, but they do not explain their existence in old and/or high metal abundance (Z$>$0.008) environments. That suggests that at least a fraction of these variable stars must have evolved via binary channels, where mass transfer or merger may play an important role \citep{bono, acep-evol}.  

\begin{figure*}
\begin{center}
\includegraphics[width=170mm]{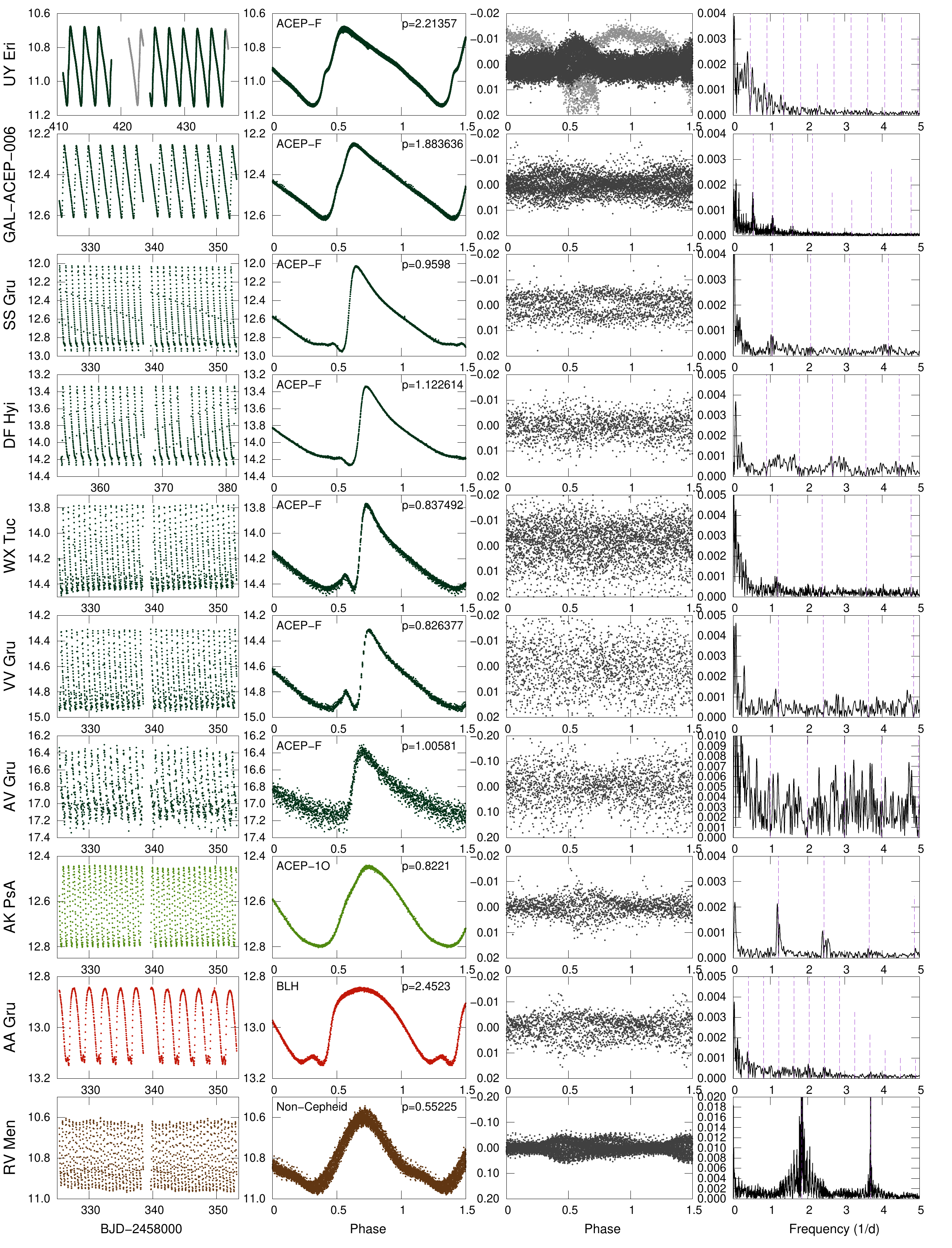}{}
\end{center}
\caption{\textit{TESS} light curves of stars classified as short-period Cepheids in the Milky Way. Types are marked with different colours. Dark green: ACEP-F, light green: ACEP-1O, red: BLH, brown: non-Cepheid. Columns from left to right: light curves, phased light curves, phased residual light curves and residual spectra (after removing the frequency series of the brightness variation, marked with dashed lines).}
\label{fig:lc1}
\end{figure*}

ACs have their own P--L relation  that differs from that of the other Cepheid classes. 
Their pulsation periods ($\sim$0.6--3 days) overlap with that of the RR~Lyrae stars, BL~Her and classical Cepheid stars. However, their distinction from the other Cepheid types is possible, since 
their light curve shapes significantly differ at a given period value. 
Despite these factors, the classification of these stars is often unreliable, mostly because of the uncertainties in absolute brightness and the lack of fine details in the light curve shape caused by noise and the sparseness of data. This is especially true for the Milky Way galaxy where, for example, a recent revision of classifications resulted in finding several new ACs previously assumed to be Type~II Cepheids \citep{jur}. Hopefully the \textit{Gaia} and \textit{TESS} surveys will solve this situation and help us to assign the proper type of variability.

ACs pulsate either in the fundamental or first overtone radial mode. The existence of the latter was revealed by the discovery of the two distinct linear P--L relationships of ACs \citep{nemec88, nemec94}, that was also confirmed by the OGLE \citep{anom-ogle, anom-ogle2} and VMC Surveys \citep{ripepi14} of the Magellanic Clouds. We now believe that nearly one third of ACs pulsate in the first overtone mode. In this paper we present the \textit{TESS} measurements of nine ACs, two of which belong to the overtone group (we marked them as ACEP-F and ACEP-1O in Table~\ref{table:targets}).

Here we briefly present the BL Her stars (BLH) as well, since our sample contains one member of this class.  BLHs are the shortest-period Type~II Cepheids (with pulsation periods between 1-5 days). They are low mass stars (0.5-0.6 M$_{\odot}$) evolving on the blue horizontal branch, and pulsating almost exclusively in the fundamental radial mode. The first double-mode pulsators among them in the bulge \citep{smolec-2018, udalski18} and two purely first overtone candidates in the LMC  \citep{sosz19} have been discovered only very recently. Theoretical calculations suggest overtone pulsation to be very rare among BLHs, in agreement with the observations.   

Fig.~\ref{fig:lc1} shows the \textit{TESS} data of ten short-period Cepheid candidates in the Milky Way we have analysed: light curves, phase curves, residual phase curves and residual Fourier spectra (after prewhitening with the main frequency and its harmonic series to the order indicated in Table~\ref{table:fou}). Our eleventh short-period Galactic target, XZ~Cet, to which we dedicate a separate section  (Sect.~\ref{sect:xzcet}), is excluded here.

In order to analyse the light curve quality, we measured the noise level by calculating the standard deviation of the residual light curves. We found it to be $\sim$4 mmag for stars brighter than 14th mag, $\sim$10 mmag for stars between 14-15th mag and $\sim$63 mmag for the faintest star, AV~Gru (16.84 mag). 
Almost all stars suffer from a slight trend that also appears as a low-frequency excess in the residual spectra. 
We suspect the origin of these trends is instrumental. In the case of WX~Tuc we identified contamination from scattered light being responsible for the amplitude change at the beginning of the \textit{TESS} orbits (BJD 2458326 and 2458340). The cause of the transient decrease in the pulsation amplitude at UY~Eri (marked with grey) is also due to instrumental effects. We discuss the various potential instrumental issues in more detail in Sect.~\ref{issues}. 

We noticed cycle-to-cycle variation on the order of 10 mmag in the only BLH of the sample, AA~Gru. This star is easily recognisable by its differing light curve shape. The same is true for the overtone anomalous Cepheid AK~PsA that has the most sinusoidal light curve in Fig.~\ref{fig:lc1}. 

The fundamental-mode ACs often have a characteristic small bump on the ascending branch of their light curves. Here such features are only noticeable in the longest period ones, UY~Eri and OGLE-GAL-ACEP-006. The latter shows remnant peaks in the residual Fourier spectrum at the pulsation frequency and its harmonics. These may originate from problems with stitching the data obtained in different sectors: contamination from neighbouring stars may differ from sector to sector that can affect the measured pulsation amplitudes. Small changes in the pulsation amplitude over time can generate side-peaks in the frequency spectrum. We experienced that with OGLE-GAL-ACEP-006, as well as at targets in the Magellanic Clouds (see Section \ref{sec:MC}.). A brighter star can be found at 76.18 arcsec separation (TYC 9368-421-1) that shows $\sim$0.4 mag variability according to \textit{Gaia} measurements, and therefore may cause the instability of the light curve of OGLE-GAL-ACEP-006. Another star, the overtone pulsator AK~PsA also shows incoherent frequencies and displays a weak amplitude modulation, similar to the Blazhko effect commonly seen in the fundamental mode RR~Lyrae stars (RRab), and less commonly in their overtone mode siblings (RRc). Unfortunately, the full modulation cycle is not covered here, if there is any. One AC with possible Blazhko modulation (or multi-mode pulsation) has been reported by \citet{stetson} in the galaxy Leo~IV, and other Blazhko AC candidates have recently been discovered from the K2 mission data by \citet{eap}, all pulsating in the fundamental mode. 

The last star in Fig.~\ref{fig:lc1}, RV Men was previously classified as first-overtone classical Cepheid in the ASAS database \citep{asas}, RRc in VSX, and as RRab by \citet{maintz} as well as by ASAS-SN \citep{asas-sn} and \textit{Gaia} DR2 \citep{Clementini}. The discrepancy of classifications of this 10th magnitude star sparked our interest. \textit{TESS} uncovered the fine details of its light curve shape, which is unusual both for Cepheids or RR~Lyraes, as well as its instability in the form of irregular phase and amplitude changes (causing a forest of remnant peaks in the residual spectra) that are unlike the Blazhko effect. Therefore, we suspect that this star is a rotational variable with stable stellar spot pattern that falsified the classification based on sparsely sampled photometric data. Stable light curves of rotational variables have recently been presented by \citet{iwanek2019}.

\subsubsection{XZ Ceti} 
\label{sect:xzcet}
\begin{figure}
\begin{center}
\includegraphics[width=85mm]{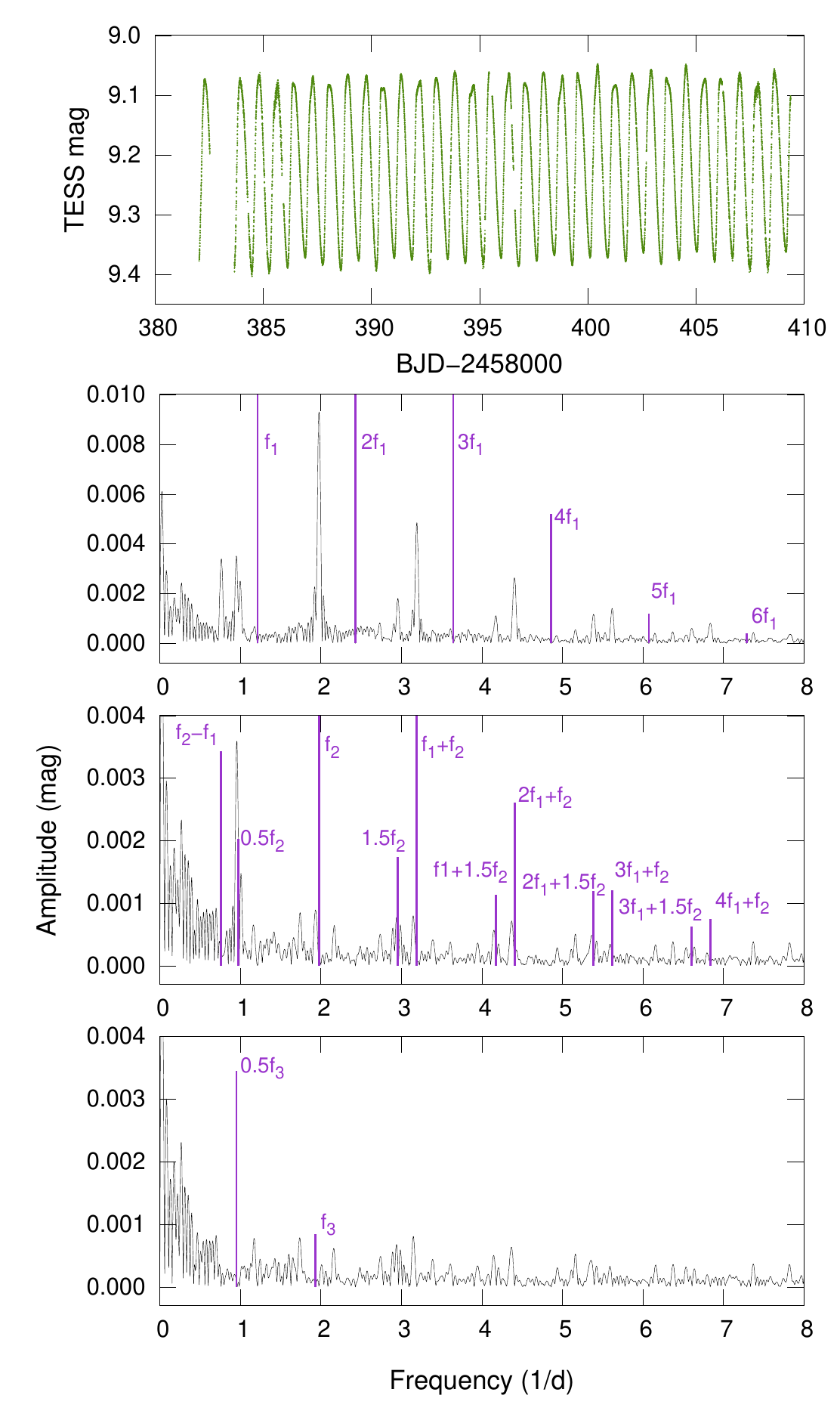}
\end{center}
\caption{The \textit{TESS} light curve of XZ~Cet (upper panel) and residual Fourier spectra at different pre-whitening steps: after pre-whitening with the $f_1$ harmonic series (second panel), with $f_2$, its subharmonics and combinations with $f_1$ (third panel), finally with $f_3$ and its subharmonic (lower panel). } 
\label{fig:xz}
\end{figure}

The variability of XZ Ceti was discovered by \citet{hoff} and the star was initially classified as RR~Lyrae type by \citet{dean}. Later, \citet{teays} fitted linear nonadiabatic pulsation models to it and concluded that XZ Ceti must be a first-overtone AC, or a first-overtone BLH.
A detailed investigation concerning the type and pulsation properties of XZ~Ceti was carried out by \citet{szabados2007}. Based on RV measurements, the authors ruled out XZ~Cet as a classical Cepheid and proposed the type AC, so XZ~Cet became the first AC identified in the Galactic field. Their analysis showed systematic shifts in the RV at certain phases, possibly indicating binarity. Instabilities in the pulsation were also reported, as well as a secular period change in the O$-$C diagram from historical data dating back to the Harvard photographic plate archive containing information from 1888 on. 

\begin{table}

\renewcommand{\thetable}{\arabic{table}}
\centering
\caption{Fourier solution of XZ Cet.} \label{tab:xzcet_fou}
\begin{tabular}{llcl}
\tablewidth{0pt}
\hline
\hline
&Frequency&Amplitude& Phase \\ 
&\hfil(c/d) & (m) & \\
\hline
$f_1$&	1.215105(8)&	 0.15298(6)& 0.13958(6)	  \\
$2f_1$&	2.43021(5)&	 0.02638(6) & 0.8554(3)	 \\
$3f_1$&	3.64532(9)&	 0.01327(6) & 0.5430(6)	 \\
$4f_1$&	4.8604(2)&	 0.00519(6) & 0.183(2)	 \\
$5f_1$&	6.0755(10)&	 0.00120(6) & 0.831(8) \\
$6f_1$&	7.291(3)&	 0.00041(6) & 0.12(2)	 \\
$7f_1$&	8.5057(10)&	 0.00092(6) & 0.734(10) \\
$8f_1$&	9.7202(10)&	 0.00104(6)	&  0.387(9)\\
$9f_1$&	10.9359(10)&	 0.00097(6)	&  0.031(10)\\
$10f_1$&	12.1511(10)&	 0.00078(6) & 0.68(12)	 \\
$11f_1$&	13.366(2)&	 0.00060(6)	&  0.34(15)\\
$12f_1$&	14.581(3)&	 0.00043(6) &	0.96(2)  \\
$13f_1$&	15.796(4)&	 0.00028(6)	&  0.61(3)\\
\hline
$f_2$&	1.97717(10)&	 0.00930(6)  & 0.8511(10)\\
$f_1+f_2$&	3.1922(2)&	 0.00462(6)	& 0.332(2)\\
$2f_1+f_2$&	4.4074(5)&	 0.00265(6)	& 0.883(3)\\
$3f_1+f_2$&	5.622(10)&	 0.00117(6)	&  0.531(8)\\
$4f_1+f_2$&	6.838(10)&	 0.00073(6)	&  0.192(12)\\
$f_2-f_1$&	0.7621(4)&	 0.00336(6) &	0.839(3)  \\
$0.5f_2$&	0.9886(3)&	 0.00353(6)  & 0.3348(11)\\
$1.5f_2$&	2.9657(7)&	 0.00152(6)   &0.819(5)\\
$f_1+1.5f_2$&	4.1808(10)&	 0.00107(6) &  0.361(10)\\
$2f_1+1.5f_2$&	5.3960(10)&	 0.00110(6) & 0.845(12)\\
$3f_1+1.5f_2$&	6.611(2)&	 0.00064(6)& 0.474(15)\\

\hline 
$f_3$&	1.9296(10)&	 0.00097(6) & 0.980(11) \\
$0.5f_3$&	0.9648(3)&	 0.00428(6) & 0.321(13)\\

\hline 
\end{tabular}
\label{table:xz}
\end{table}

\begin{figure}
\begin{center}
\end{center}
\includegraphics[width=82mm]{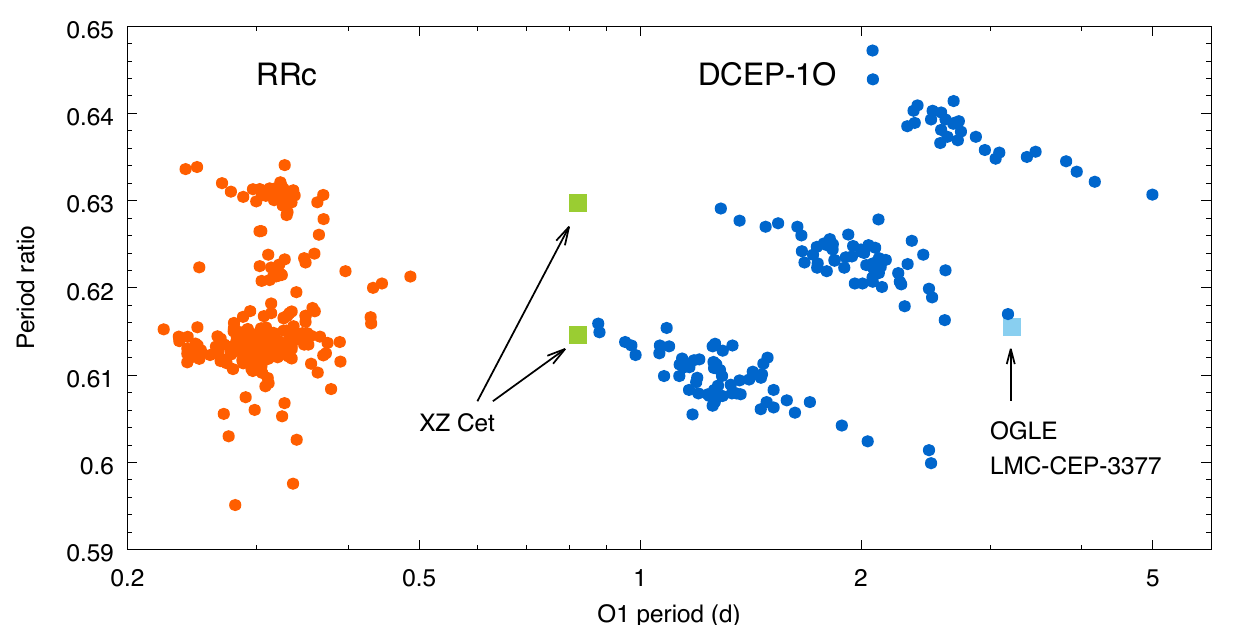}
\caption{The Petersen diagram of non-radial modes of spherical-harmonic degree $\ell = 7, 8, 9$, from top to bottom, in overtone RR~Lyrae (orange) stars and Cepheids (blue). Sources: \citet{netzel-2015,netzel2-2015,netzel-2019,moskalik15,jurcsik-2015,smolec-snieg, moskalik-2009}. The new discoveries, XZ Cet and OGLE~LMC-CEP-3377 (see Sect. \ref{sec:add}) are marked with squares.}
\label{fig:peter}
\end{figure}

The short-cadence \textit{TESS} light curve of XZ~Cet uncovered further interesting features in this star (Fig.~\ref{fig:xz}). After prewhitening with the main frequency and its 13 harmonics, a group of low-amplitude frequencies appeared, of which we could clearly identify the $\sim$0.61 mode ($f_2$) and its linear combination with the overtone mode (Table~\ref{table:xz}). We found 12 other frequencies above our 4.0 S/N amplitude threshold. In the past, no additional modes were reported in any anomalous Cepheid yet, let alone this many. We discovered a familiar pattern in the frequencies, the subharmonic structure connected to $f_2$, observed previously in \textit{Kepler} RRc stars \citep{moskalik15}. Between 0.5$f_2$ and 1.5$f_2$, the former has higher amplitude while the latter has combinations with the main pulsation frequency ($f_1$+$1.5f_2$, 2$f_1$+1.5$f_2$, $3f_1$+$1.5f_2$). The 0.5$f_2$ subharmonic was also detected in OGLE overtone RR~Lyrae and classical Cepheid stars that show $f_2$, but mostly in the form of power excess rather than a single peak \citep{smolec-snieg,netzel-2019}. 
Further analysis of the low-amplitude peaks uncovered a frequency ($f_3$) with $\sim$0.63 period ratio, and its higher-amplitude subharmonic ($0.5f_3$). 
We note, however, that the separation between 0.5$f_2$ and 0.5$f_3$ is smaller than the frequency resolution ($\sim$0.0365~${\rm d}^{-1}$) and for $f_2$ and $f_3$ is just above that. So we treat $f_3$ with caution. 
Low-amplitude oscillation near 0.61$P_1$ was first found in the RRd star, AQ~Leo, measured by the \textit{MOST} satellite \citep{grub-2007}. Since then it has become evident that a small fraction ($\sim$8-9\%) of overtone pulsators exhibit weakly excited modes with 0.60–0.64 period ratio, as this has been shown for RR~Lyrae stars in the Galactic bulge \citep{netzel-2019} and classical Cepheids in the SMC \citep{sos2010}. In addition, significant fraction (39\% for the SMC and 47\% for the LMC) of overtone Cepheids of the Magellanic Clouds show power excess with these period ratios \citep{snieg-smolec-2018}.

\begin{figure}
\begin{center}
\includegraphics[width=85mm]{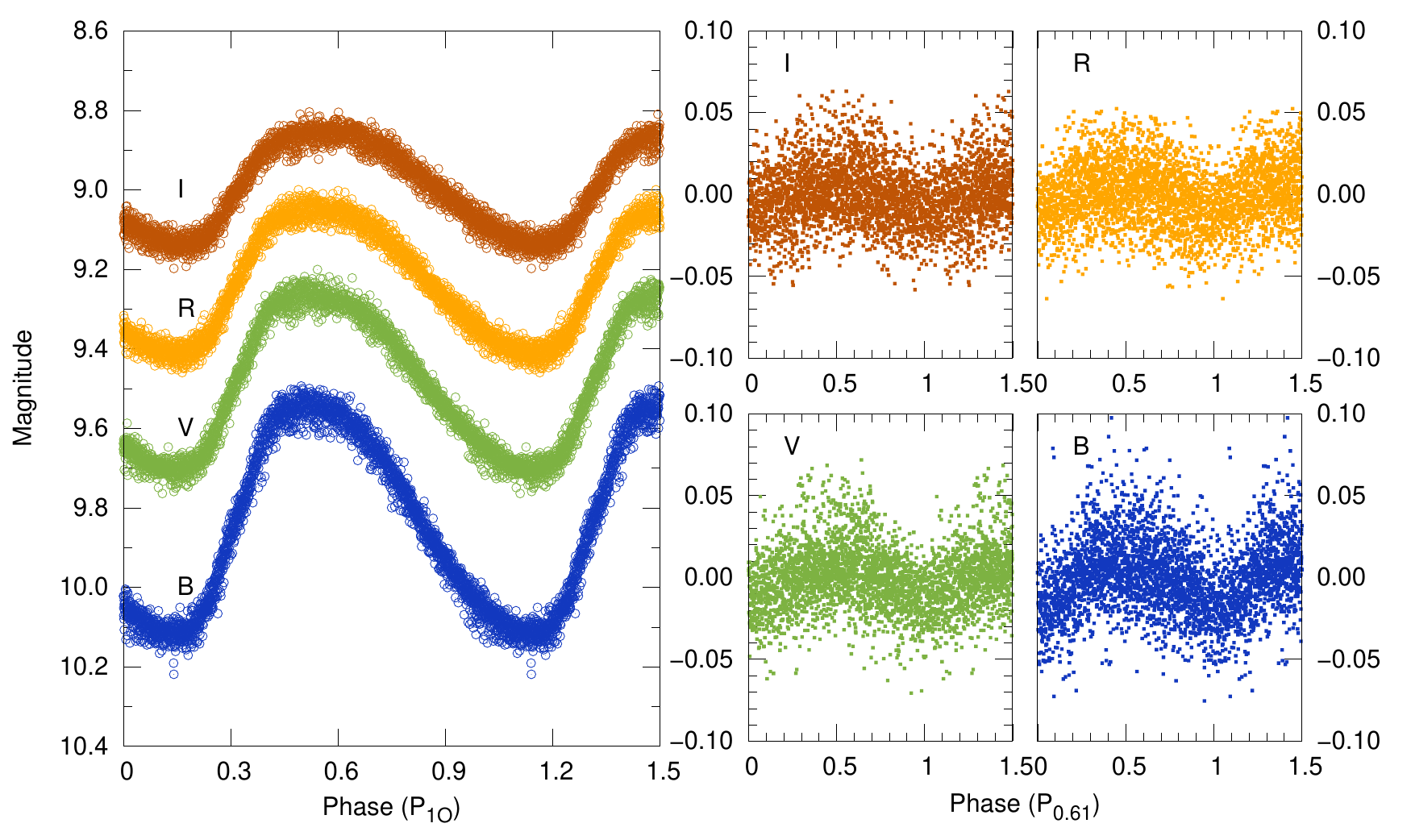}
\end{center}
\caption{Ground-based measurements of XZ~Cet in \textit{BVRI} bands. Left: phase curves folded with $P_\mathrm{1O}$=0.82315 days.
Right: residual phase curves folded with $P_{0.61}$=0.50479 days.}
\label{fig:hmb}
\end{figure}

\begin{table}

\renewcommand{\thetable}{\arabic{table}}
\centering
\caption{Fourier amplitudes for XZ Cet light curves in various photometric bands.} \label{tab:xzcet_tab}
\begin{tabular}{lccc}
\tablewidth{0pt}
\hline
\hline
Filter &$A_\mathrm{1O}$& $A_{0.61}$& $A_{0.61}/A_\mathrm{1O}$\\ 
\hline

I &	0.1413(5) &	 0.0077(5)	&	0.054(4) \\
R &	 0.1831(6) &	0.0086(6)  &		0.047(3) \\
V &	 0.2227(6)&	 0.0114(6)	&	0.052(3) \\		
B &	 0.2921(6)&	0.0128(6)  	&	0.044(2) \\

\hline 
\textit{TESS} &	 0.15298(6)&	0.00930(6)  	&	0.0608(4) \\
\hline

\end{tabular}
\label{table:hmb}
\end{table}

\begin{figure*}
\begin{center}
\includegraphics[width=160mm]{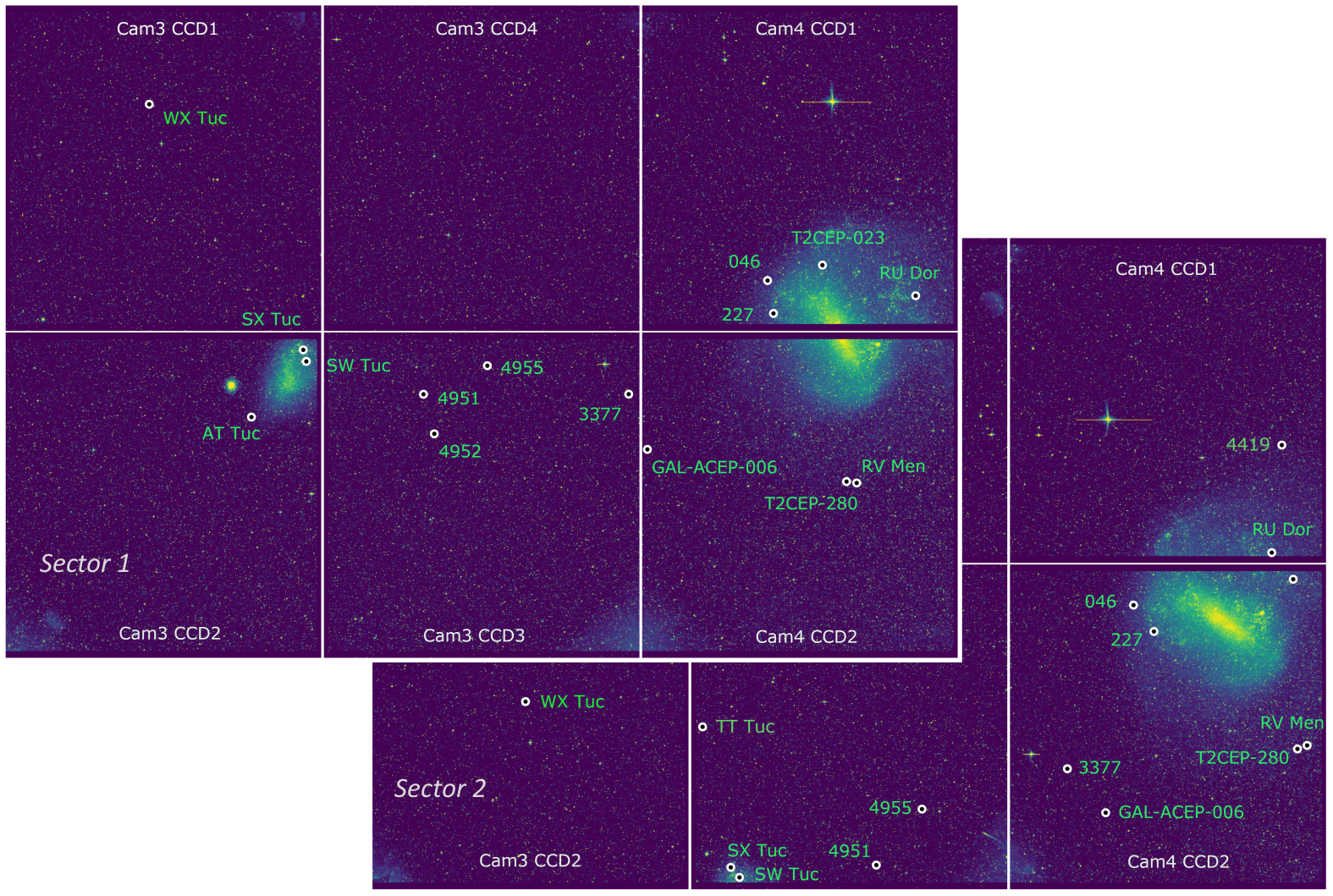}
\end{center}
\caption{Position of targets in and near the Magellanic System in Sectors 1 and 2.}
\label{fig:ccd}
\end{figure*}
These modes, that are sometimes referred to as $f_x$ (or $f_{0.61}$) modes, are hypothesised to be non-radial modes \citep{dziem}, where the actual pulsation frequency is the 0.5$f_x$ peak that typically appears with smaller amplitude (or does not appear at all) due to geometric cancellation.
We note, however, that it can appear with larger amplitude, or is the only peak detected (i.e. harmonics are not detected, \citealt{smolec-snieg,suveges-2018,netzel-2019}).
The non-radial modes of spherical-harmonic degree $\ell=7$, 8, and 9 form sequences in the Petersen diagram (from top to bottom, respectively), that slightly differ for RR~Lyrae and classical Cepheids stars (Fig. \ref{fig:peter}). In RR~Lyrae stars only the $\ell=8$ and $\ell=9$ modes seem to be present with a smaller middle sequence caused by their combination peaks. The $f_2$ and  $f_3$ modes of XZ~Cet are placed just between the variability classes in period, and clearly match the $\ell=8$ and $\ell=9$ sequences.

New multicolour photometric observations were obtained for XZ~Cet at the Remote Observatory Atacama Desert (ROAD, \citealt{road}) in Chile between 16 October 2019 and 13 January 2020. Data are available in  the AAVSO database (with HMB observer code). We have analysed the data taken in \textit{BVRI} passbands and discovered that $f_2$ mode is visible in all four bands, and the $f_1$+$f_2$ combination, too, in the \textit{B} band. We plotted the phase curves and the residual phase curves after removing the main periodicity with four harmonics and folded with $P_{0.61}$ (see Fig.~\ref{fig:hmb}). We also calculated the Fourier amplitudes and ratios in different colours (Table \ref{table:hmb}). The extensive study of RRc stars in the globular cluster M3 by \citet{jurcsik-2015} reported colour measurements of the 0.61 mode for the first time. The authors showed that the RRc stars with the 0.61 mode are bluer than the regular ones and redder than the RRd stars, being exactly between these two groups in the $(B-V)$~-~$(V-I)$ plane.

The \textit{TESS} light curve of XZ~Cet provided the first detection of non-radial pulsation in an AC type star, and hopefully more will follow in the future.

\begin{figure*}
\begin{center}
\includegraphics[width=180mm]{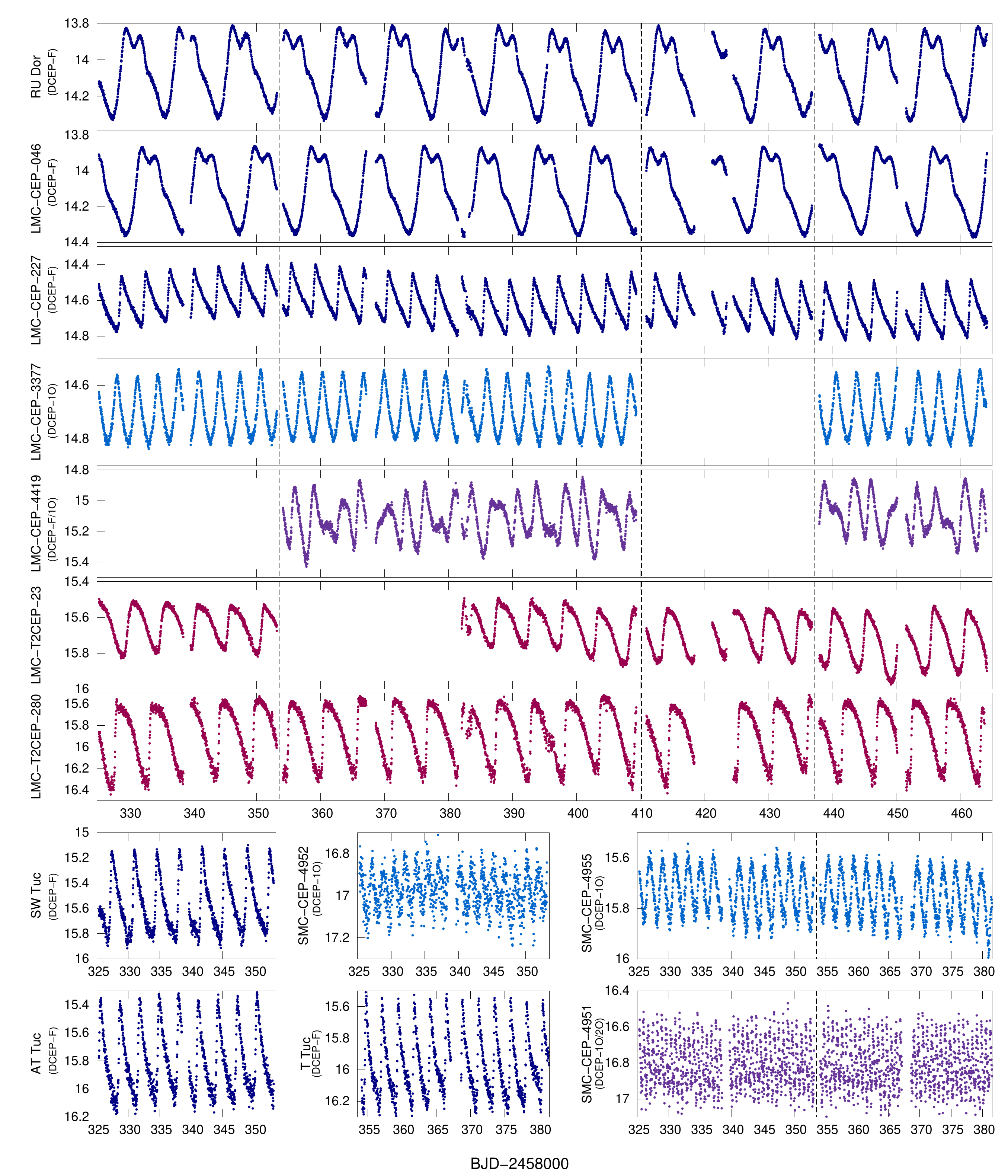}{}
\end{center}
\caption{FITSH light curves of Cepheids in the Magellanic System. Subtypes are marked in different colours: DCEP-F (dark blue), DCEP-1O (light blue), double-mode stars (purple), T2CEP (dark red). Dashed vertical lines divide the different sectors. }
\label{fig:Maglc}
\end{figure*}

\subsection{Cepheids in the Magellanic System} \label{sec:MC}

\textit{TESS} observed both Magellanic Clouds in Sectors 1 and 2. Moreover, the LMC was continuously observed during the entire first year of the mission. It is our misfortune, however, that as a consequence of the P--L relation, the shortest period Cepheids are the faintest ones, seriously limiting our target selection in the Magellanic System. To avoid large confusion, we selected our targets at lower-density stellar fields in the outskirts of the LMC and the SMC, and from the Magellanic Bridge. We also added two stars from the inner region of SMC to test the limits of \textit{TESS}. Targets in \textit{TESS} CCD frames of Sector 1 and 2 are displayed in Fig.~\ref{fig:ccd}.
This region of the sky has been extensively studied by the OGLE Survey, thus we have the opportunity to compare all of our findings with OGLE data. The FITSH light curves of the selected Cepheids in the Magellanic System are plotted in Fig.~\ref{fig:Maglc}.

\subsubsection{Bump Cepheids}
We chose two bump Cepheids with similar periods: RU~Dor and OGLE~LMC-CEP-046 (Fig. \ref{fig:Maglc}). The bump feature appears on the descending branch of the light curves of fundamental-mode classical Cepheid (DCEP-F) stars with pulsation periods between 6 to 10 days, and moves backwards in phase as the period increases: this phenomenon is known as the Hertzsprung progression. (The bump appears on the ascending branch between periods 10 to 20 days.) 
Both stars show cycle-to-cycle variation, with up to 40 mmag in the case of OGLE~LMC-CEP-046 and 60 mmag for RU~Dor. This range is above the precision limit of OGLE measurements, but no indication of such amplitude fluctuation is visible in those data. Therefore we suspect that the observed cycle-to-cycle variations are partially or even entirely of instrumental origin. The observed pulsation periods have not changed since the OGLE measurements, they are identical at the two epochs within the errors. 

Because of the presence of a bump on the light curve, RU~Dor was marked earlier as a probable Population~II star \citep{rudor}. Indeed, bump Cepheids can be mistakenly classified as W~Vir-type stars due to the similarities in light curve shape and period. Fortunately, an extensive classification was carried out in the OGLE Survey based on Wesenheit indices \citep{madore1982} where these two types can be reliably distinguished in the Magellanic Systems \citep{anom-ogle, sos2010}. With these studies at hand our knowledge on light curve differences has expanded considerably.

\subsubsection{pWVir stars}

\begin{figure}
\begin{center}
\includegraphics[width=85mm]{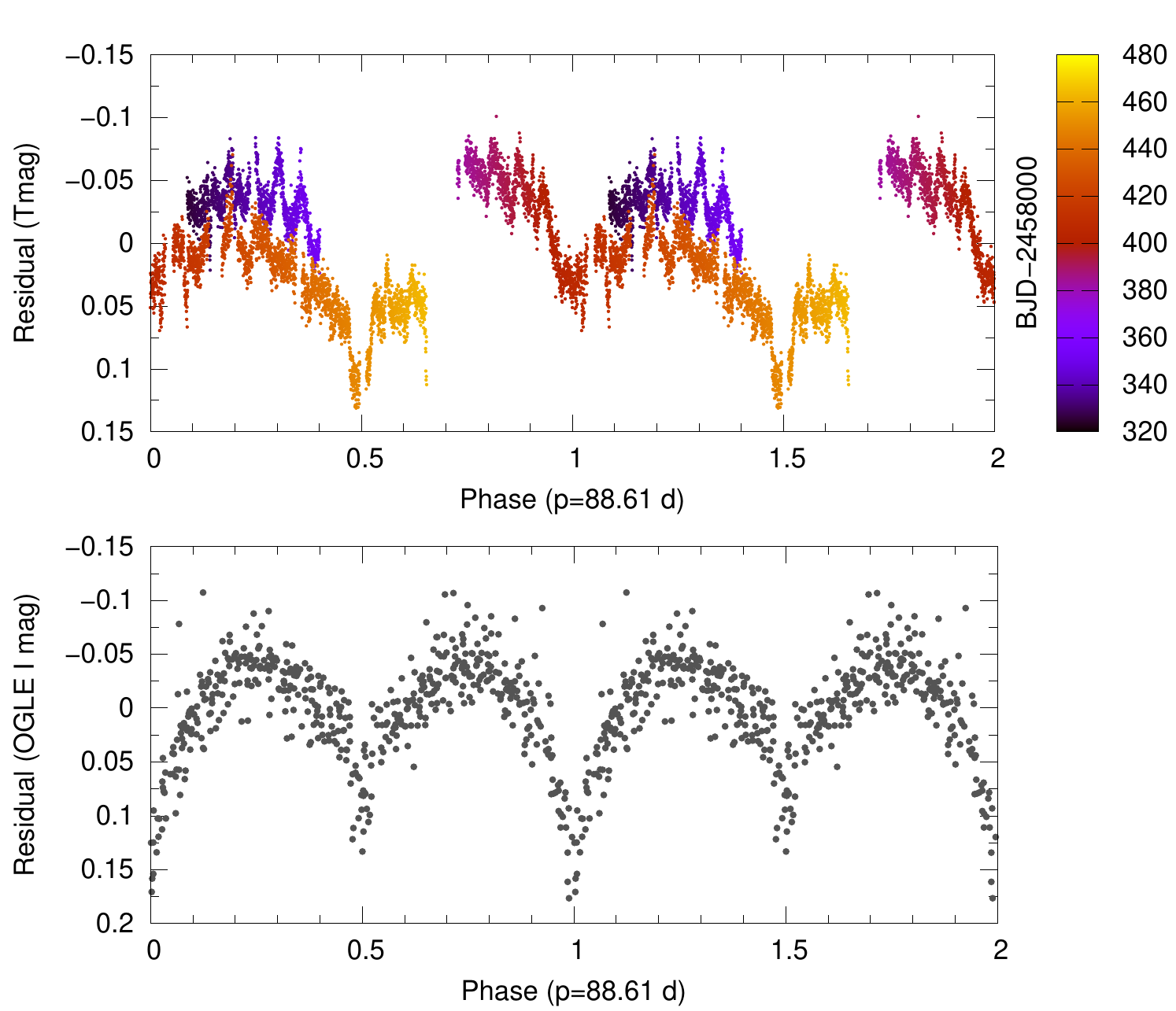}
\end{center}
\caption{Residual light curves of OGLE LMC-T2CEP-023 folded with the orbital period of the system. Upper panel: \textit{TESS} data. Colour bar indicates the time evolution (BJD-2458000). Lower panel: OGLE I-band data.}
\label{fig:23}
\end{figure} 
W Vir stars represent the intermediate-period subtype of Type~II Cepheids: they are 1.5–2 mag fainter than the classical Cepheids at similar pulsation periods, making their \textit{TESS} photometry more challenging. Here we present two members from the so-called `peculiar W~Vir' group (OGLE LMC-T2CEP-023 and OGLE LMC-T2CEP-280, Fig.~\ref{fig:Maglc}). These stars are on average brighter and bluer than the regular ones, and can also be recognised from light curve morphology, since shapes are closer to those of the DCEP-F type. Only 34 members of this group are known in the Magellanic Clouds \citep{pekuli} and the same amount in the Galactic bulge \citep{acep-bulge1}. Half of the group show signs of binarity in the form of ellipsoidal or eclipsing variation superimposed on the pulsation. The same was detected in OGLE LMC-T2CEP-023, where an 88.61~day orbital period could be determined \citep{sos2010}. The \textit{TESS} light curve of this star clearly shows the eclipses beside cycle-to-cycle changes of the pulsation and a long-period additional variability. We removed the pulsation signal and plotted the residual light curves folded with the orbital period for both the \textit{TESS} and OGLE data in Fig.~\ref{fig:23}. These residual light curves reveal ellipsoidal variability due to tidal distortion effects, as well as a difference between the widths of eclipses. The latter phenomenon could be an indication of a disk being present in the system, suggesting that mass transfer was happening in the past, as in the case of OGLE-LMC-T2CEP-211, a similar eclipsing system with a pWVir star analysed by \citet{pilecki2018b}. We also found the combination frequencies of $1/P_\mathrm{puls}+1/P_\mathrm{orbit}$  and $1/P_\mathrm{puls}-2/P_\mathrm{orbit}$, the latter of which is also seen in the OGLE data. The combination frequencies are evidence for phase modulation caused by binary motion, and they are often detected in Type~II Cepheid light curves if the Cepheid is a member of an eclipsing or an ellipsoidal binary system.

OGLE~LMC-T2CEP-023 is a strongly blended star, with several variables within and nearby the aperture, with  periodicities known from OGLE Survey. We found excess power near to the pulsation frequency and harmonics that may originate from the neighbouring variable star OGLE~LMC-CEP-0615, which has a pulsation period of $\sim$5.1 d, very close to the 5.23 d period of T2CEP-023. We could not identify the origin of the mean brightness decrease: it might be of instrumental origin or an intrinsic feature of the system.



The other pWVir star, OGLE LMC-T2CEP-280 is located in a less dense stellar environment, away from the center of the LMC. We observe cycle-to-cycle changes in this star with amplitudes up to 10 mmag, which was not seen in the OGLE data. The scatter of the light curve is also significant ($\sim$5 mmag) and the time scale of the sectors causes a prominent peak in the Fourier spectrum highlighting the difficulty of accurate data stitching, which is especially challenging when cycle-to-cycle variations occur. Due to the instrumental issues, we cannot be certain that the observed cycle-to-cycle variation is intrinsic, although it is characteristic of Type~II Cepheids.  

\subsubsection{OGLE LMC-CEP-227}
\label{sec:227}
OGLE LMC-CEP-227 is the famous Cepheid variable in a double-lined eclipsing binary system, which was the first classical Cepheid star with a precisely determined mass \citep{ogle2008,natur,pilecki2018}. Binary systems with a Cepheid component are of great importance because dynamical masses and radii can be directly determined and compared to pulsation masses and radii, providing a link between theoretical models of stellar evolution, atmospheres, and pulsation. OGLE LMC-CEP-227 has been the subject of many detailed studies (e.g. \citealt{moroni, marconi13}) including various arguments concerning the p-factor, the projection factor that relates observed RVs to intrinsic pulsation velocities \citep{pilecki13}. The \textit{TESS} data of this star have the potential to further increase the precision of the derived stellar properties and the p-factor itself.

\begin{figure}
\begin{center}
\includegraphics[width=85mm]{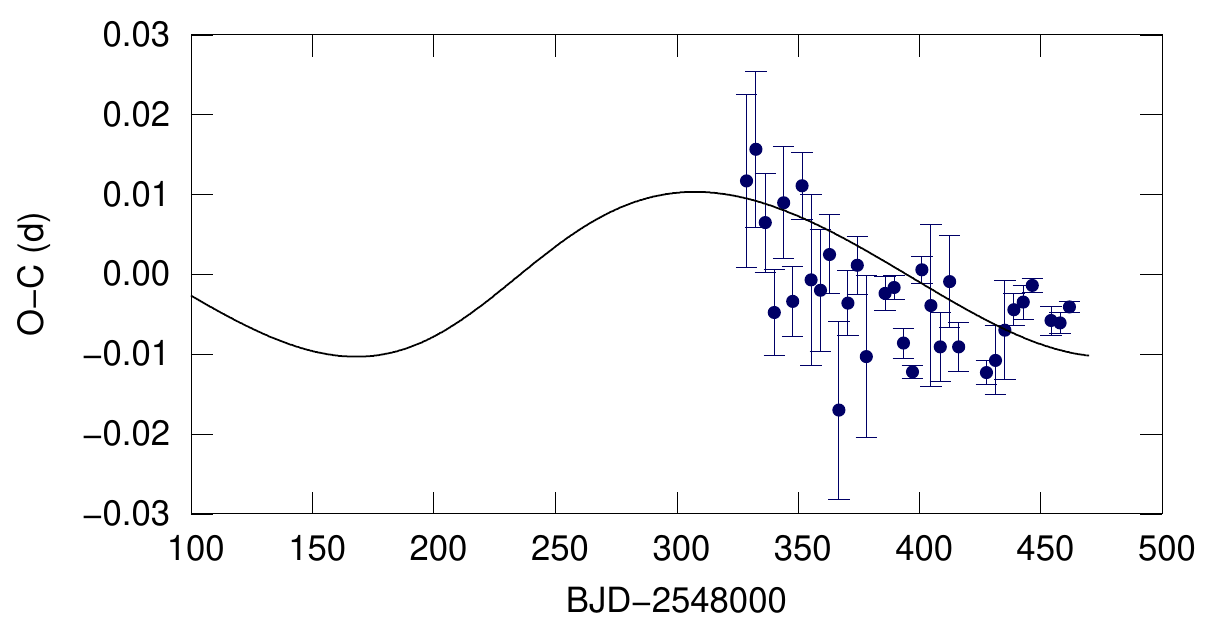}
\end{center}
\caption{The O$-$C diagram of OGLE LMC-CEP-227 plotted on the expected light-time effect caused by the known companion star.}
\label{fig:227}
\end{figure}

Here we focus only on the 
detectability of the light-time effect. OGLE LMC-CEP-227 is the primary component of the system with an orbital period of 309.404$\pm$0.002 d. In Sectors 1-5 the stars were in the orbital phase when the Cepheid component was approaching us but did not reach the eclipse. We calculated the light-time effect with the parameters given by \citet{pilecki13}, and we constructed the O$-$C diagram, displayed in Fig.~\ref{fig:227}, where `O' is defined by 3rd-order polynomial fits to individual maxima.
The pulsation period was determined to be the best fit to the light-time effect: $C$=JD2454896.285+3.7970782$E$. The O$-$C follows the slope of the curve but it is doubtful if we could recognise the light-time effect only from that. The success depends on the amplitude of the light-time effect, the duration of observational interval and the uncertainty of O$-$C residuals. The latter is related to both the brightness and crowding. 

New calculations based on \textit{Gaia} DR2 put the binary fraction of Galactic Cepheids above 80\% \citep{kervella-2019}. This is a significantly higher fraction than the 29$\pm$8\% value found by \citet{evans-binary} based on radial velocity data, although that was considered as a lower limit by the authors too. A list of binary Cepheid systems is maintained by \citet{szabados-2003}\footnote{\url{https://konkoly.hu/CEP/intro.html}}.
On the whole, we have good chances to discover binarity if we can separate light-time effect from evolutionary period changes in the O$-$C data. Such analysis would benefit enormously from repeated visits of the continuous viewing zones over multiple mission extensions. 
In contrast, we have the possibility to catch eclipses, as we do in the case of OGLE LMC-CEP-227 in the later \textit{TESS} sectors. However, light-time effect is not sensitive to all types of binaries, as demonstrated by \citet{groenewegen-2017} who could not identify the known binary OGLE LMC-T2CEP-023 this way.

We note that we did not use any detrending algorithm on any of our light curves to demonstrate the effect of contamination. In the case of OGLE LMC-CEP-227 we notice a long term variation in the mean brightness, not observed in the OGLE data  (Fig.~\ref{fig:Maglc}). After investigating the possible blend stars, we concluded that OGLE-LMC-LPV-5076 is the most probable source of this trend.

\subsubsection{Detectability of additional modes}
\label{sec:add}
We mentioned in Sect.~\ref{sect:xzcet} that non-radial modes may be excited in overtone pulsators. Here we report a case where the non-radial mode was not detected in OGLE, but is clearly seen in the \textit{TESS} data. OGLE LMC-CEP-3377, located in the Magellanic Bridge, was identified in OGLE-IV \citep{bridge15} and was classified as first-overtone classical Cepheid (DCEP-1O). The additional mode we found in the \textit{TESS} light curve belongs to the middle sequence of $f_x$ modes (Fig.~\ref{fig:peter}). The amplitude of this mode is 2 mmag, which is below the detection limit of OGLE-IV (3 mmag).
It is very likely that this frequency comes from the Cepheid star, as there are no known variables in its close neighbourhood. However, we observed an additional slow, sinusoidal variation in the light curve (with $\sim$20 mmag full amplitude) that may originate either from a blended source, from a companion star, or from instrumental effects. O$-$C calculations were not helpful in this case, more data and investigations are needed.

\begin{table*}[ht]
\renewcommand{\thetable}{\arabic{table}}
\centering
\caption{Frequency content and Fourier amplitudes of OGLE LMC-CEP-4419 light curves constructed with 2.5, 2.0, 1.5, and 1.0 pixel radius apertures respectively. }
\label{tab:4419}
\begin{tabular}{llcccc}
\tablewidth{0pt}
\hline
\hline
ID &Frequency & A(2.5\,px)& A(2.0\,px)& A(1.5\,px)& A(1.0\,px)  \\ 
\hline

$f_0$&      		 	0.28744(4)  & 	 0.1030(8)	&	0.0828(7)	&	0.0745(5) 	&	0.0546(4)\\
$f_1$&       			0.40093(4)  &    0.0972(8) 	&	0.0802(7)	&	0.0711(5) 	&	0.0514(4)\\
$f_2$=$2f_0$--$f_1$&   	0.17613(6)  & 	 0.0658(8)   	&	0.0249(7)	&	0.0109(5) 	&	--\\
$f_0$+$f_1$&  	  		0.68837(10) &	 0.0346(8)  	&	0.0287(7)	&	0.0249(5) 	&	0.0178(4)\\
$2f_2$=$4f_0-2f_1$    &      0.35225(11) &    0.0312(8) 	&	0.0092(7)	&	--		&	--\\	
$f_1-f_0$&     		0.1135(2)   &    0.0192(8)  	&	0.0192(7)	&	0.0165(5) 	&	0.0119(4)\\
$2f_0$&    	  		0.5749(2)   &	 0.0186(8)  	&	0.0154(7)	&	0.0136(5)	&	 0.0100(4)\\
$3f_2$=$6f_0$--$3f_1$   &	0.5284(3)   &    0.0125(8)    	&	--		&	-- 	&	 --\\
$2f_0$+$f_1$ &    		0.9758(4)   &    0.0103(8)    	&	0.0075(7)	&	-- 	&	--\\

\hline
\end{tabular}

\end{table*}

\begin{figure}
\begin{center}
\includegraphics[width=80mm]{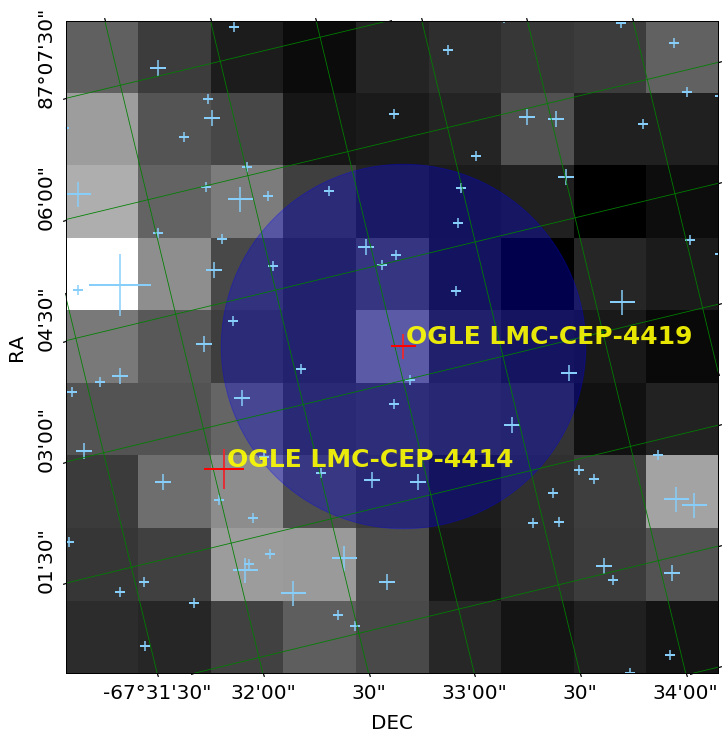}
\end{center}
\caption{Image cutout of OGLE LMC-CEP-4419. The 2.5 pixel radius aperture is marked in blue. Crosses are stellar objects identified in \textit{Gaia} DR2.}
\label{fig:4419}
\end{figure} 

We have to be careful to interpret signals as extra modes in Cepheids. Here we describe another interesting case of a spurious additional frequency found in the double mode Cepheid OGLE LMC-CEP-4419.
Table~\ref{tab:4419} summarises the results of our Fourier analysis. We detected the fundamental mode ($f_0$, 2$f_0$) and the overtone mode ($f_1$),  as well as their linear combinations  ($f_0$+$f_1$, $f_1$--$f_0$, $2f_0$+$f_1$). In addition, we detected a low-amplitude frequency series that can be explained either as combination frequencies: $2f_0$--$f_1$, $4f_0$--$2f_1$, $6f_0$--$3f_1$ or as a third mode $f_2$, $2f_2$, $3f_2$. But these combinations would be very unusual, and $f_2$ would not fit to the known frequency ratio groups in the Petersen diagram as an additional mode \citep{suveges-2018}. Therefore we investigated the neighbouring stars and found that $f_2$ is the pulsation frequency of the fundamental-mode classical Cepheid, OGLE LMC-CEP-4414, located 63.37$^{\prime\prime}$ away (Fig.~\ref{fig:4419}). To check the contamination from this star we created light curves with smaller apertures using radii of 2, 1.5 and 1 pixels. The smaller the radius, the weaker the contamination, as it is visible in Table~\ref{tab:4419}, and it drops below the detection limit at the 1.0 px aperture.
Nevertheless, we do not consider the tight apertures to be better in general, even if we can avoid a large fraction of the contaminations. The intrinsic low-amplitude modes may also remain undetectable if only a small fraction of the flux is collected.

\subsubsection{Cepheids in the SMC}

\begin{figure*}
\begin{center}
\begin{minipage}[t]{.45\linewidth}
\includegraphics[width=79mm]{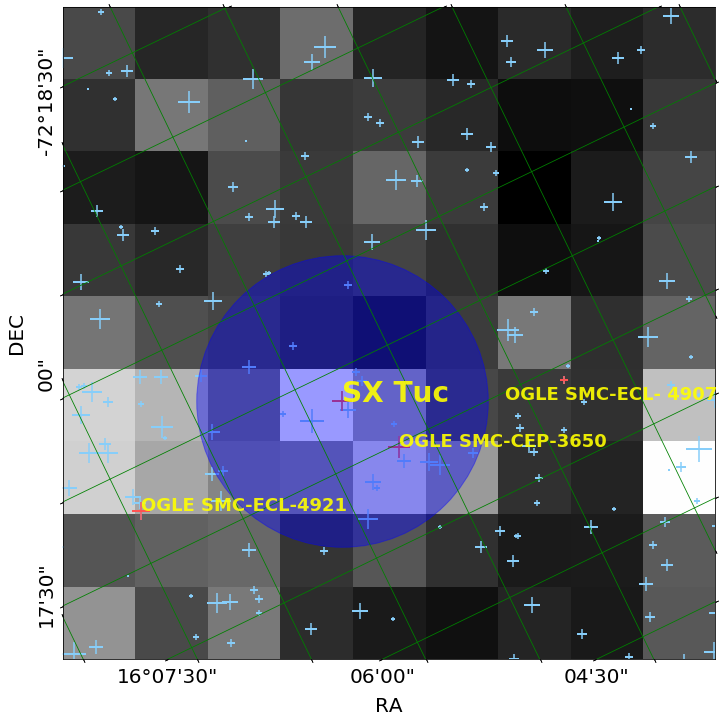}
\end{minipage}
\begin{minipage}[b]{.45\linewidth}
\includegraphics[width=88mm]{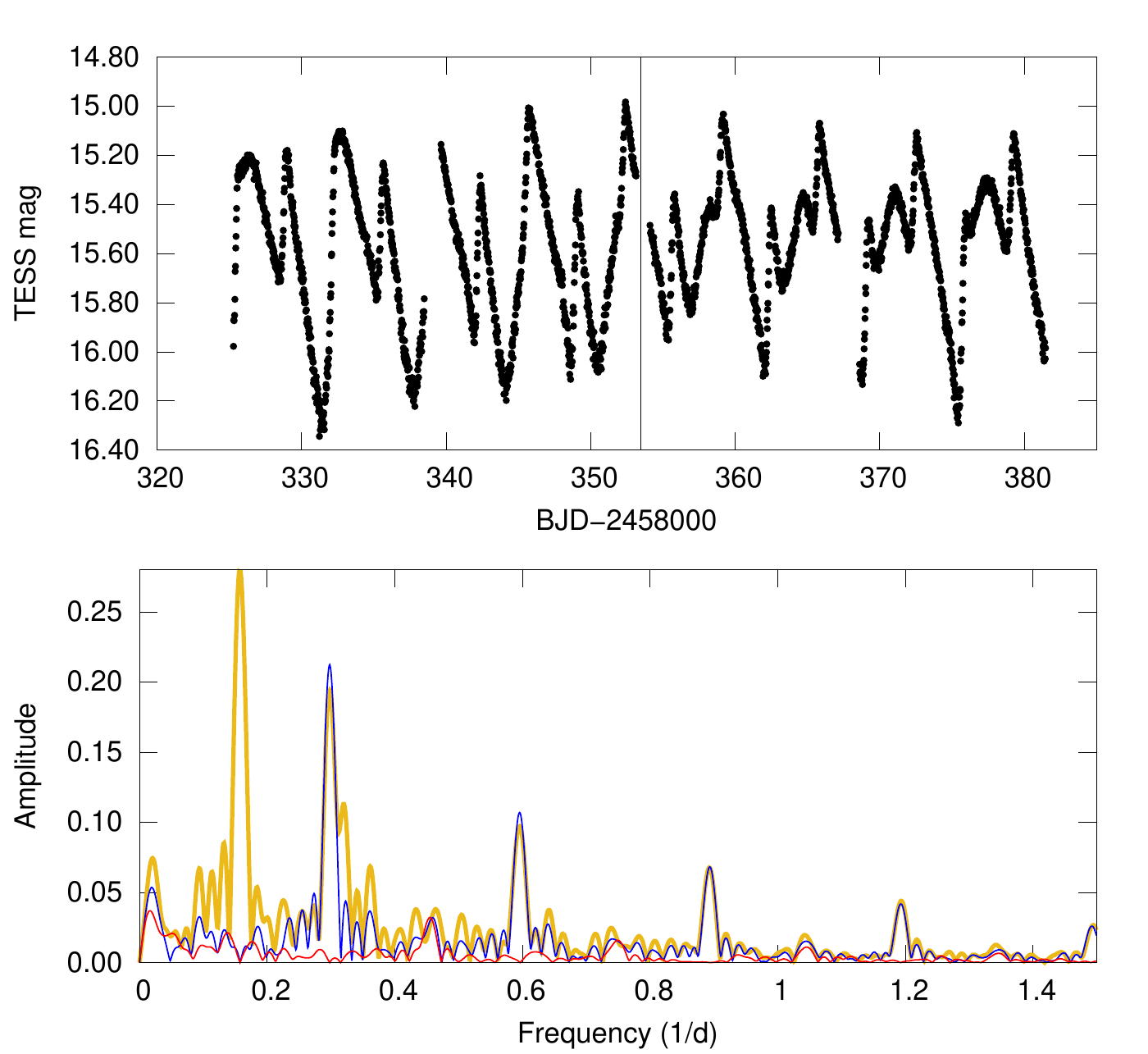}
\end{minipage}
\end{center}
\caption{Left: image cutout of SX Tuc and OGLE SMC-CEP-3650 within the aperture. Upper right: composite light curve. Lower right: composite Fourier spectrum (yellow), SX~Tuc in blue, residual in red.)}
\label{fig:sxtuc}
\end{figure*}

We identified another case of strong contamination between Cepheids in the SMC as well. Our DCEP-F type target star, SX~Tuc, lies away from the dense centre of the galaxy, but its surroundings are still crowded with variable stars. Even with 2-px apertures we could only generate a light curve that is a composite that also contains variation from a neighbouring Cepheid at 21.41$^{\prime\prime}$ distance, OGLE SMC-CEP-3650. This latter star is brighter and its variability dominates the composite Fourier spectrum displayed in Fig.~\ref{fig:sxtuc}. Moreover, the first harmonic frequency of OGLE SMC-CEP-3650 is very close to the pulsation frequency of SX~Tuc. Despite all these complications, we were able to determine the harmonic series of SX~Tuc up to the seventh harmonic. The Fourier parameters were compared with the OGLE values, and we found only a few percent difference. This implies that contamination does not make Fourier parameter calculations necessarily impossible, but may cause some uncertainties. 

We examined three DCEP-F stars in the SMC (see Figs.~\ref{fig:ccd} and \ref{fig:Maglc}). SW~Tuc shows a changing light curve shape. We found frequencies in the residual spectra at $\sim$0.87, $\sim$1.74, and $\sim$0.62 c/d. Although, there are no known variables in the neighbourhood, contamination is the most likely explanation for these signals. TT~Tuc and AT~Tuc are located in the outskirts of the SMC. In the light curve of TT~Tuc we detected the pulsation frequency of the neighbour OGLE-SMC-RRLYR-1868 at $\sim$1.42 c/d. AT~Tuc displays cycle-to-cycle changes, possibly caused by contamination from an unidentified source. 

We selected two DCEP-1O type stars from the Magellanic Bridge: OGLE SMC-CEP-4952 \citep{bridge16} and OGLE SMC-CEP-4955 \citep{bridge15}.
The former is the farthest known Cepheid in the Magellanic Bridge, and the faintest star in our sample. Its type is also problematic, since it has recently been reclassified as DCEP-F based on distance calculations using optical period–Wesenheit relations \citep{bridgenew}. However, the $R_{21}$ and $R_{31}$ Fourier parameters based on the OGLE light curves do not support the reclassification. We could reliably detect the first harmonic from the \textit{TESS} light curve, but there is no clear peak at the second harmonic frequency above the noise level. Therefore we can only give an upper limit for the $R_{31}$ value, at $R_{31}<0.065$. That puts this star very far from the DCEP-F domain in the $\log P-R_{31}$ plane, but agrees with the region where classical overtone and Type~II Cepheids are located (Fig.~\ref{fig:fouparams}). The proper classification of OGLE SMC-CEP-4952, as well as its origin in that far out region of the Magellanic Bridge remains uncertain. 

The overtone OGLE SMC-CEP-4955 is a blended star, which might be responsible for the trend visible in its light curve. This trend also shows the periodicity of the \textit{TESS} orbits. No additional mode is detected beyond the main frequency and its two harmonics.

Finally, we present the only known double-mode Cepheid in the Magellanic Bridge, OGLE-SMC-CEP-4951. Both the O1 and O2 modes could be detected in \textit{TESS} data, the former along with two harmonics, and one combination frequency is also visible. The orbital period of \textit{TESS} causes weak peaks at $\sim$0.07 and $\sim$0.14  c/d.

\begin{figure*}[ht]
\begin{center}
\includegraphics[width=0.95\textwidth]{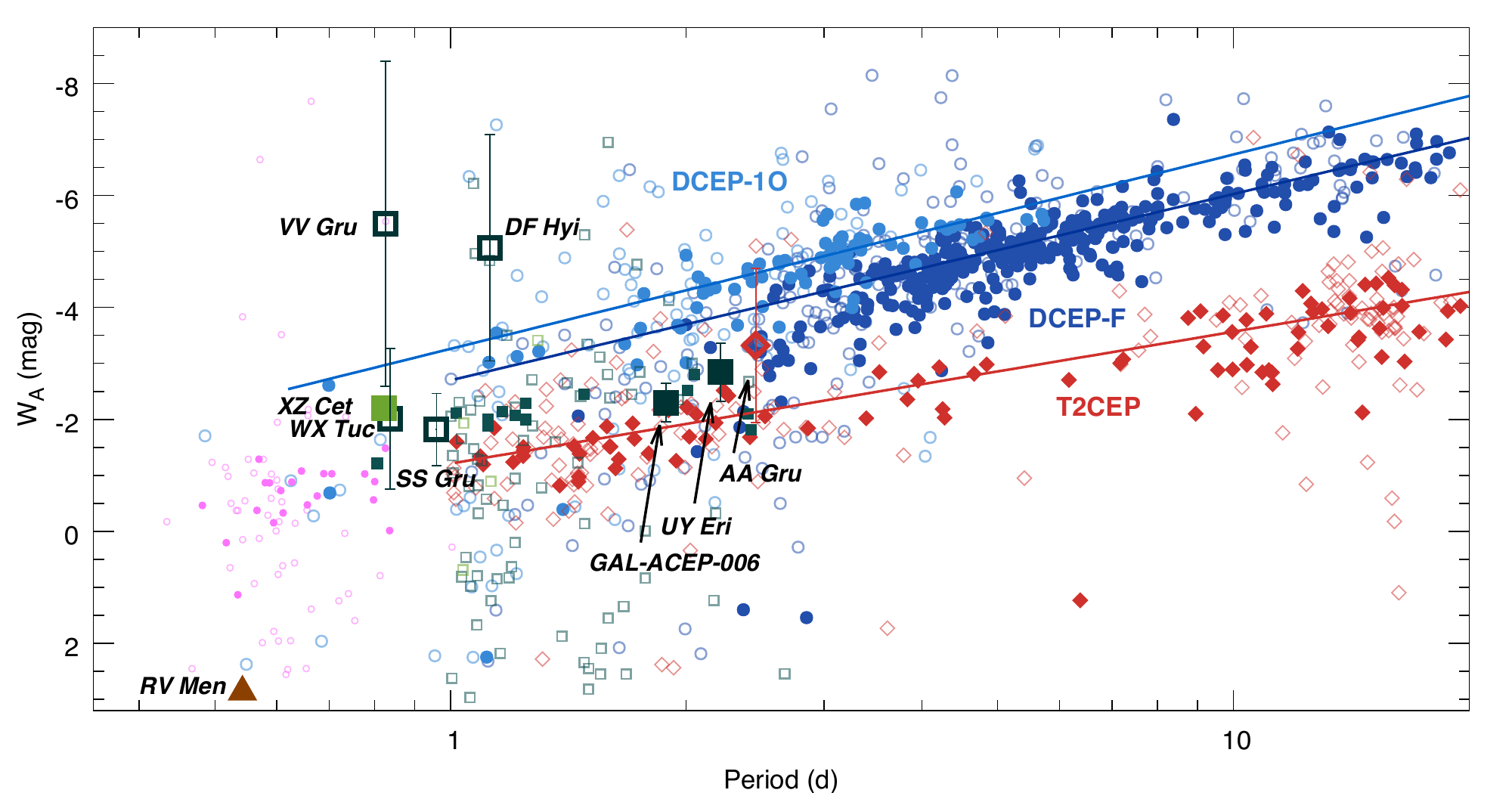}
\end{center}
\caption{P-L relation of Milky Way Cepheids using absolute Wesenheit magnitudes \citep{vincenzo}. Blue dots: DCEP-F, light blue circles: DCEP-1O, green squares: ACEP-F, light green squares: ACEP-1O, red diamonds: T2CEP, pink: stars reclassified as RRab. Filled symbols indicate stars with parallax errors below 20\%. Our target stars are indicated with large symbols. The non-Cepheid RV Men is the brown triangle at the short-period end. Lines: P-L relations calculated by \citet{vincenzo} for each class, slopes have been fixed to LMC Cepheids.}
\label{fig:plr}
\end{figure*}

\begin{figure*}
\begin{center}
\includegraphics[width=150mm]{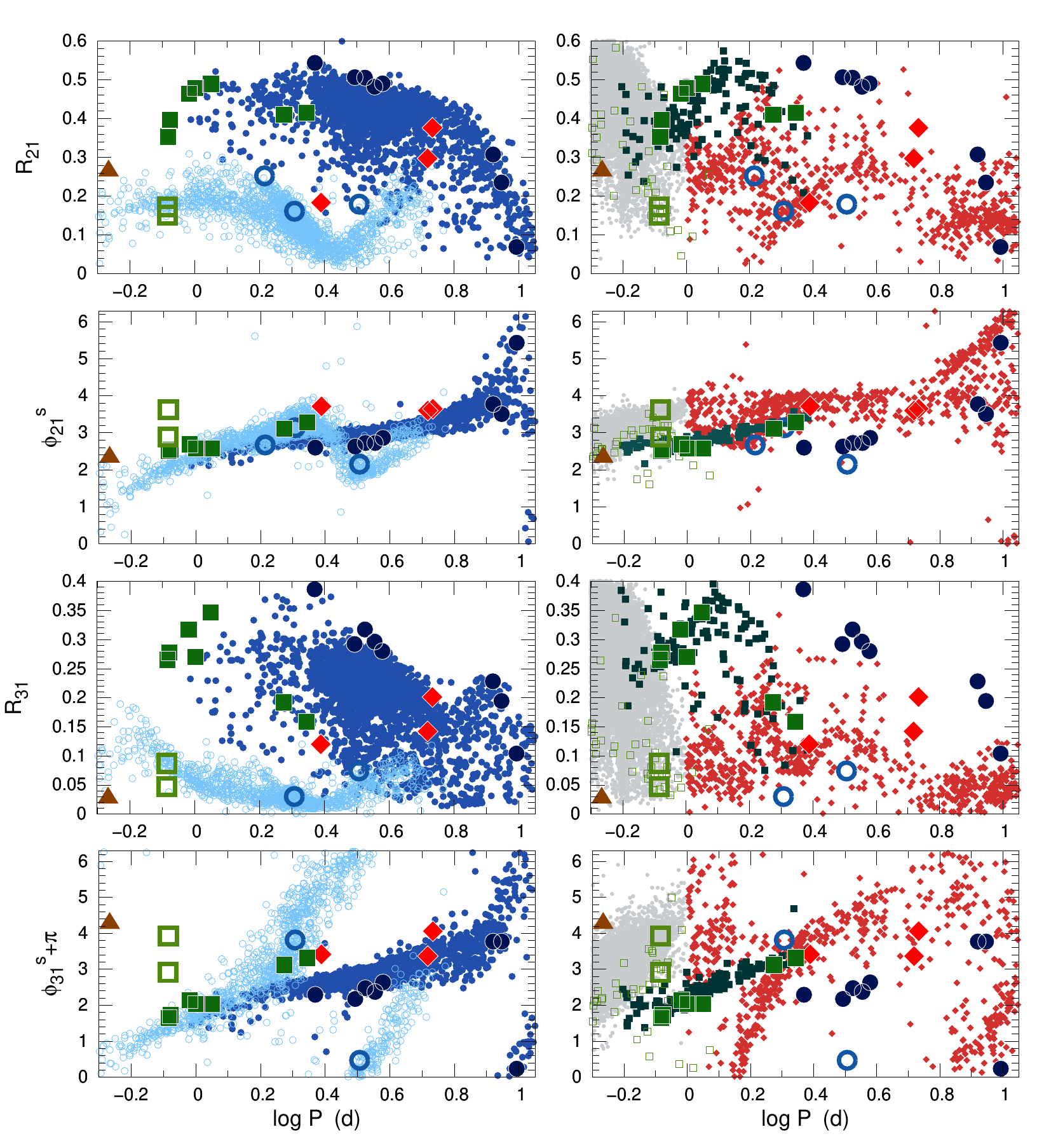}
\end{center}
\caption{Relative Fourier parameters $R_{21}$, $\phi_{21}$,  $R_{31}$ and $\phi_{31}$ against the pulsation periods. $S$ indices indicate that  cosine-based phase values provided in OGLE were converted to sine-based values used in Table \ref{tab:results}: ${\phi_{21}}^{ S}$=$\phi^C_{21}-\pi/2$, ${\phi_{31}}^S$=$\phi_{31}^C+\pi$. Large symbols are our targets with colours corresponding to the most likely classes (colours are the same as in Fig. \ref{fig:plr}.). Small symbols are the OGLE-IV Fourier parameters for Classical Cepheids (left panels), and Type II and anomalous Cepheids (right panels). We also included RRab stars with grey dots in the latter.}
\label{fig:fouparams}
\end{figure*}

\subsubsection{On the instrumental issues}
\label{issues}

We demonstrated that even if the angular resolution of \textit{TESS} is suboptimal for studying dense fields, useful science can still be obtained with careful analysis. It is essential to investigate the surroundings of the target stars, but some phenomena may remain unexplored even after doing so. We already mentioned the possibility of discovering eclipses. Given the large fraction of Cepheids in binary systems, it is possible that eclipses show up in the continuous \textit{TESS} observations. We have not detected any eclipses in our program stars yet, but we faced an instrumental issue that can mimic an eclipse. This temporary fading is visible in all light curves at the beginning of Sector 3, best seen in both pWVir stars. The origin of this anomaly is a test of the Attitude Control System (ACS), which was conducted several times during Sector 3, but affected the data mostly at this part. Thanks to these tests an improved ACS algorithm was developed and installed that reduced pointing jitter in the later sectors. Near the end of the orbits in Sector 3 stray light features caused by the Earth distorted the light curves. There are multiple types of data anomalies that include different types of scattered light and optical artefacts. These are collected and described in detail in the \textit{TESS} Data Release Notes\footnote{https://archive.stsci.edu/tess/tess\_drn.html} and the \textit{TESS} Instrument Handbook \citep{handbook}. In the final analysis we excluded those  sections of the light curves that are strongly affected by the instrumental issues.  

\subsection{Classification and Fourier parameters} \label{sec:foupar}

Cepheids are one of the primary distance indicators in the local Universe \citep{hubble-riess-2018}, therefore the accuracy of our distance measurements can depend on our ability to identify them properly. The OGLE Catalogue is a good source of classifications for the Magellanic System, where both the light curve shape and the brightness information are used. The latter is applied through a reddening-free brightness quantity, the Wesenheit magnitude. However, the accuracy of Cepheid distances may also depend on blending corrections for remote targets in crowded fields, as pointed out by \citet{majaess-2020}.

The identification of Cepheids in the Galactic field is more uncertain. Except for the brightest, well studied Cepheids, we detect a significant fraction of mis-classifications among Cepheid candidates \citep{szabo-2011,poretti-2015}.
A considerable part of the Cepheid candidates identified in \textit{Gaia} DR2 had low absolute magnitudes, well below the expected P-L relation, indicating that they are either other variables or their parallaxes are inaccurate \citep{Clementini}. These were subsequently re-classified in a follow-up work that produced a cleaner sample \citep{vincenzo}. The Galactic Cepheids, in particular, were analysed in detail, employing both numerical parameters and visual inspections of the light curves along with period-Fourier parameter and period-absolute Wesenheit diagrams plus literature sources. 

We adopted the same formulae used in that paper to calculate the \textit{Gaia} Wesenheit magnitudes $(W = G-1.90\,(G_{\rm BP}-G_{\rm RP}))$ and calculated the absolute Wesenheit magnitudes of those \textit{TESS} stars whose offset-corrected parallaxes were above zero $(W_A = W + 5\,\log \varpi_{\rm corr}+5)$. We used the same 0.046 milliarcsecond (mas) parallax zero-point offset calculated by \citet{Riess2018} for consistency. We note that the \textit{Gaia} DR2 parallax zero-point offset was estimated by various authors and published values span wide range from 0.0 to --0.1 mas (e.g., \citealt{groen-2018,stassun-2018,marconi-2020}). Corrected parallaxes remained negative for AV~Gru and AK~PsA. Due to the lack of $G_{\rm BP}$ and $G_{\rm RP}$ values we could not calculate $W_A$ for $\beta$~Dor either. We plotted the original and cleaned DR2 Cepheid samples and the P-L relations obtained from the cleaned DR2 sample in Fig.~\ref{fig:plr}. We then overplotted our \textit{TESS} Milky Way stars to see if they line up with the PLRs or the loci defined by the \textit{Gaia} classifications. As Fig.~\ref{fig:plr} demonstrates, the absolute Wesenheit magnitudes are too uncertain for these stars for a conclusive result, indicating that the underlying parallaxes themselves are too uncertain (see Table \ref{tab:par}). This is not unexpected: \citet{vincenzo} themselves state that especially at the short-period end of the Cepheid populations, the different classes overlap. Even with the Fourier parameters of the \textit{Gaia} light curves at hand, they could not classify all objects unambiguously, concluding that collecting well-sampled, high-precision light curves remains the best way to solve this issue.   

\begin{table}
\renewcommand{\thetable}{\arabic{table}}
\centering
\caption{\textit{Gaia} DR2 parallaxes and uncertainties of our Milky Way subsample used in Fig \ref{fig:plr}, before correcting for the zero-point shift.} 
\label{tab:par}
\begin{tabular}{lcc}
\hline
\hline
Name & $\varpi$& $\sigma_{\varpi}$\\
&(mas)& (mas)\\
\hline
UY Eri &  0.248 & 0.057    \\
OGLE GAL-ACEP-006 &	0.134	&	0.021	\\	
SS Gru	&	0.138	&	0.039	\\
DF Hyi	&	-0.023	&	0.022	\\		
WX Tuc	&	0.037	&	0.021	\\	
VV Gru	&	-0.033	&	0.044	\\		
XZ Cet		&		0.793		&		0.046		\\		AA Gru		&		0.041		&		0.026		\\		RV Men		&		4.560		&		0.036		\\	

\hline
\end{tabular}
\end{table}

Unambiguous classification of pulsating stars based exclusively on light curve shape and period would be a significant achievement in stellar astrophysics. There are many clear cases, where reliability of the classification depends exclusively on the quality of the photometric data. 
It was demonstrated nearly forty years ago that the light curve structure in its quantitative form can describe the Hertzsprung progression \citep{simon-lee-1981} and it is a sensitive tool to discriminate the pulsation modes in RR~Lyrae stars \citep{simon-teays-1982} and classical Cepheids. Since then Fourier decomposition has become a widely used classifier for pulsating stars. However, as with all techniques, it has its own limitations that mostly arise from close similarities between the light curves of pulsators and other variable stars (rotational variables and eclipsing/ellipsoidal binaries). The diverse light curve shapes that appear within the subtypes of Cepheids complicate the problem even more.

The Fourier parameters R$_{21}$, R$_{31}$, $\phi_{21}$ and $\phi_{31}$ form clusters for the various types of Cepheids when plotted against the pulsation period. In Fig.~\ref{fig:fouparams} we plotted the OGLE-IV $I$-band values for LMC and Galactic bulge Cepheids \citep{ogle-blg-rab-2014,bridge15,anom-ogle2,acep-bulge1,pekuli}, where we separated the classical Classical Cepheids (left panels) from Type~II and anomalous Cepheids (right panels) for better visibility. Since the passband of \textit{TESS} is similar to the OGLE $I$ band, we could overlay the Fourier parameters of our target stars listed in Table \ref{table:fou}. Fourier parameters calculated from the OGLE data are available for many of our targets, and found them to differ from \textit{TESS} values only by a few percent, which changes the position of the stars on the plots only slightly. The clusters in Fig.~\ref{fig:fouparams} seemingly overlap on the individual plots, but if we take all four plots into account, class membership becomes more clear. 
We note, however, that the two peculiar W Vir stars (marked with large red diamonds at $\log P\sim$0.7) cannot be distinguished from DCEP-F stars based only on their Fourier parameters.
The Cepheid impostor, RV Men is also indicated (with brown colour). Based on its location, this star can belong to the RR~Lyrae stars, demonstrating the uncertainty of this classifier method, in which the Fourier coefficients represent average values for the light curves and no fast changes are taken into account. 

Many Galactic Cepheid candidates observed by large sky surveys have under-sampled light curves from which Fourier parameters can be determined only with large errors. The precise and continuous light curves of \textit{TESS} will certainly solve this problem and availability of fine details of the light curve structure allows us to filter out the non-pulsating variables. 
We note, however, that the low-order Fourier parameters are only simplified quantitative measures of the average light curve and this fact is independent of the photometric accuracy. Using more information embedded in the fine details of brightness variation, can lead to significant progress in stellar classification, and this way we could utilise the \textit{TESS} precision more. A recent attempt in this direction has been made by \citet{szklenar-2020} who used phased light curves of diverse variable stars to test an image-based classification attained with machine learning methods. \textit{TESS} may provide a new basis for such efforts. 

\subsubsection{Comparison with models}
\label{sec:masses}
The light curve shape of RR~Lyrae and Cepheids stars depend on global stellar parameters. The pioneering study of \citet{jurcsik-kovacs-1996} presented a formula for RRab metallicity as a linear function of the period and the $\phi_{31}$ coefficient. An analogous correlation has been found for the RRc stars as well \citep{morgan-2007}. Such a clear relation between the light curve shape and the metallicity is not known for Cepheid stars. Although there are indications that metallicity affects the pulsation amplitudes of Cepheids, the nature of the dependence has been debated \citep{szabados-klagyivik-2012,majaess-2013}. Recent efforts to estimate fundamental parameters such as mass, radius, luminosity, and effective temperature from photometry also demonstrated that light curve structure is statistically important both in RR~Lyrae and Cepheid types \citep{bellinger-2020}. 

The efficacy of the Fourier decomposition technique to compare observed and theoretical Cepheid light curves has been recognised and widely used since the 1980s \citep{simon-davis-1983, petersen-hansen-1984, stellingwerf-1986}. 
A recent comparative study explores the variation of Fourier parameters at different compositions and mass-luminosity levels as a function of the wavelength and the period \citep{anupam-2017} to provide stringent constraints for the stellar pulsation models (\citealt{bono2000,bono2010,marconi13} and references therein). 

\begin{figure}
\begin{center}
\includegraphics[width=90mm]{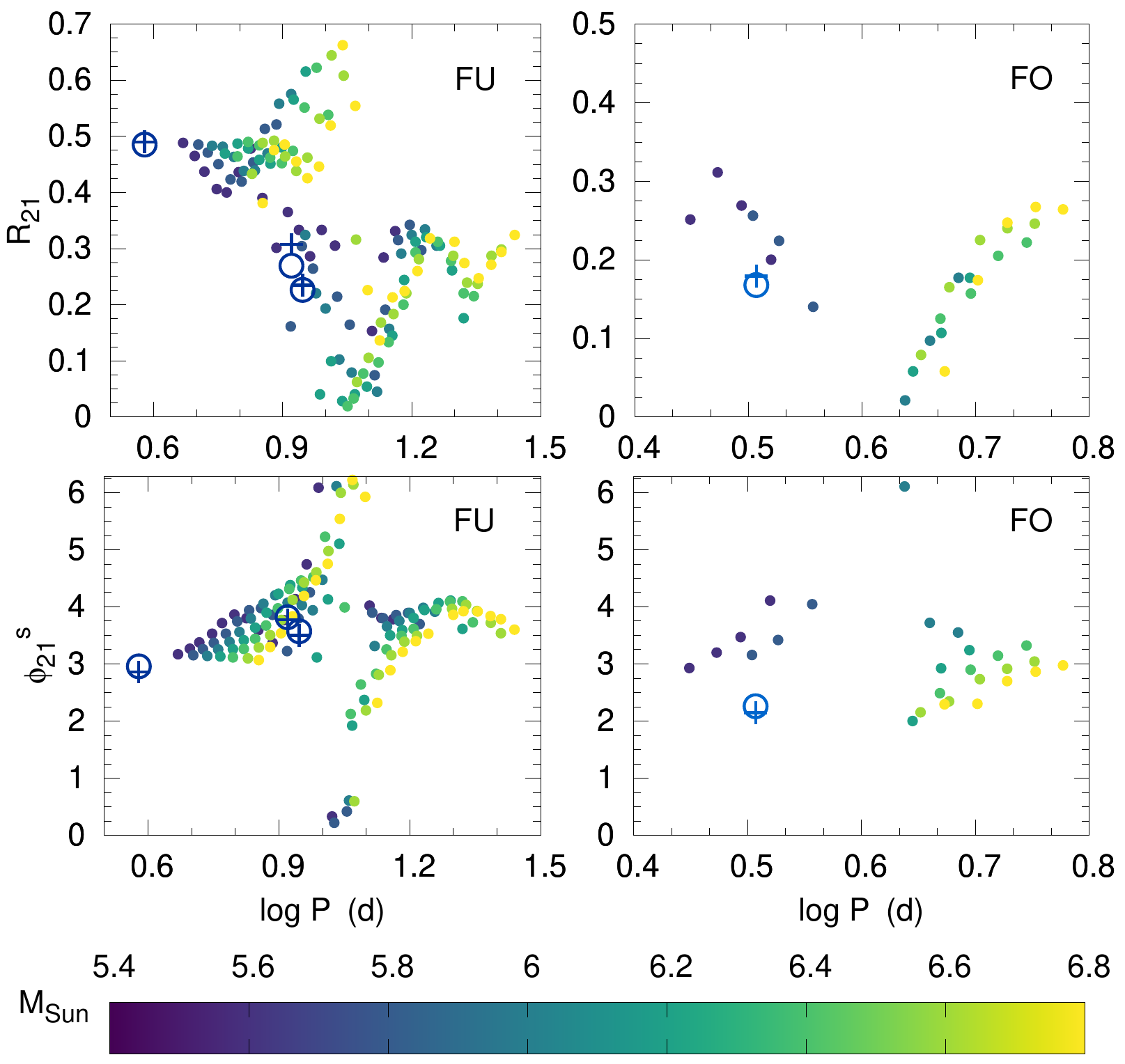}
\end{center}
\caption{Theoretical Fourier parameters for LMC Cepheids at different masses, for LMC-like chemical composition (Y=0.25, Z=0.008). 
Observed \textit{TESS} (plus) and OGLE I-band (circle) Fourier parameters of OGLE LMC-CEP-227, RU Dor, OGLE LMC-CEP-046 and OGLE LMC-CEP-3377 are indicated.}
\label{fig:mass}
\end{figure}

We adopted the \textit{I}-band models from \citet{anupam-2017}, computed with metallicities representative for the LMC (Y=0.25, Z=0.008), as described by \citet{marconi13}, and compared them with our parameters. Our sample contains only three fundamental and one overtone Cepheids whose Fourier parameters can be derived from both the \textit{TESS} data and OGLE \textit{I}-band photometry and we can directly compare them to those of the models in Fig.~\ref{fig:mass}, for example as a function of stellar mass. The offset between the parameters of the two surveys is in the range of model grid resolution, thus precise fit can be obtained only after  
recomputing models and generating light curves in the \textit{TESS} passband. Extending the grid toward shorter periods, where \textit{TESS} observations are the most efficient, would also be welcome. Given the models covering a range of chemical composition, the combination of \textit{TESS} and OGLE observations and theoretical models may be the basis of promising future investigations on physical parameters of Cepheids in the Galaxy and the Magellanic Clouds.   
 
\subsubsection{High-order Fourier parameters}

\begin{figure}
\begin{center}
\includegraphics[width=85mm]{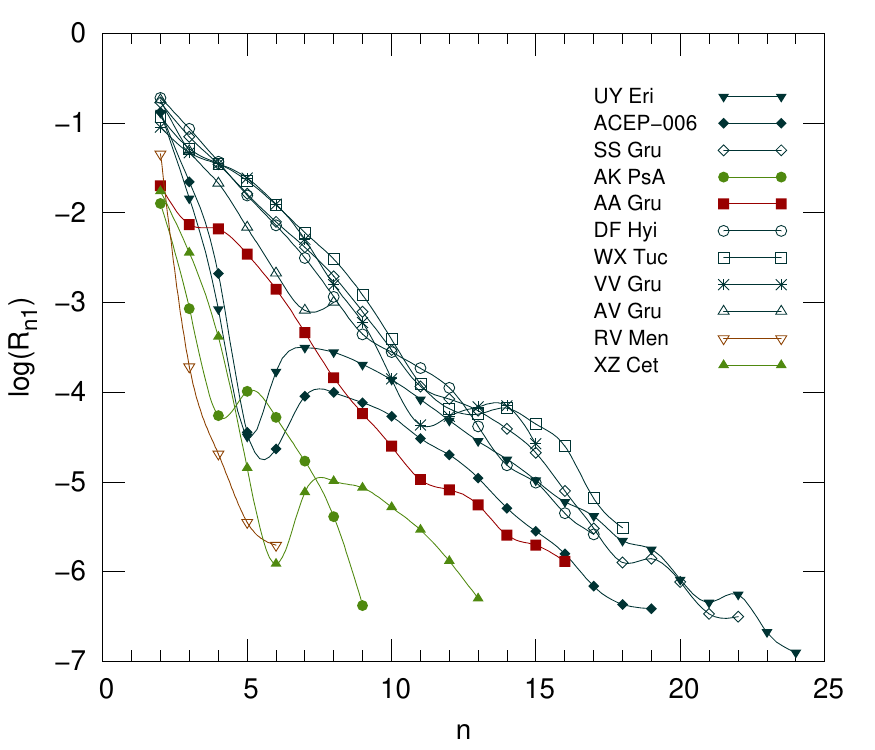}
\end{center}
\caption{Amplitude ratios vs. harmonic order ($n$) for the short period Cepheids: ACEP-F1 (blue), ACEP-O1 (green), BLH (dark red), and RV Men, the impostor star (orange).}
\label{fig:harm}
\end{figure}
Although the light curve shape can be approximately described with the Fourier parameters of the main frequency and its first two harmonics, further harmonics may provide considerably more details. The number of detectable harmonics is a function of data sampling (cf.~the Nyquist frequency) and the noise level. With space-based photometric missions, high number of harmonics (10-20) could be detected for a large number of Galactic field pulsating stars, providing a basis for statistical studies of high-order harmonic structures that could be useful for stellar classification, too. We plotted the high order Fourier parameters for the amplitude ($R_{n1}$) in Fig. \ref{fig:harm}. A similar plot was presented for some \textit{CoRoT} RRab stars by \citet{benko-2016} who discovered that a deep local minimum exists in most amplitude ratio curves, and the position of the minimum is dependent on the pulsation period. 

Here we have only a small sample, but we can still declare that for the four ACEP-F stars that show significant minima in the amplitude ratio series, the longer the period, the lower the harmonic order where the minimum occurs. Moreover, both overtone stars show minima but at different harmonic orders, despite their periods being almost equal. The impostor star, RV~Men has the steepest slope. The BLH star has a similar $R_{21}$ value as the overtone ACs, but then goes on a different path. We intend to analyse more \textit{TESS} light curves to figure out what these characteristics can be used for.                                                                                                       
\subsection{Timing and O$-$C techniques} \label{sec:inst}

\begin{figure*}
\begin{center}
\includegraphics[width=160mm]{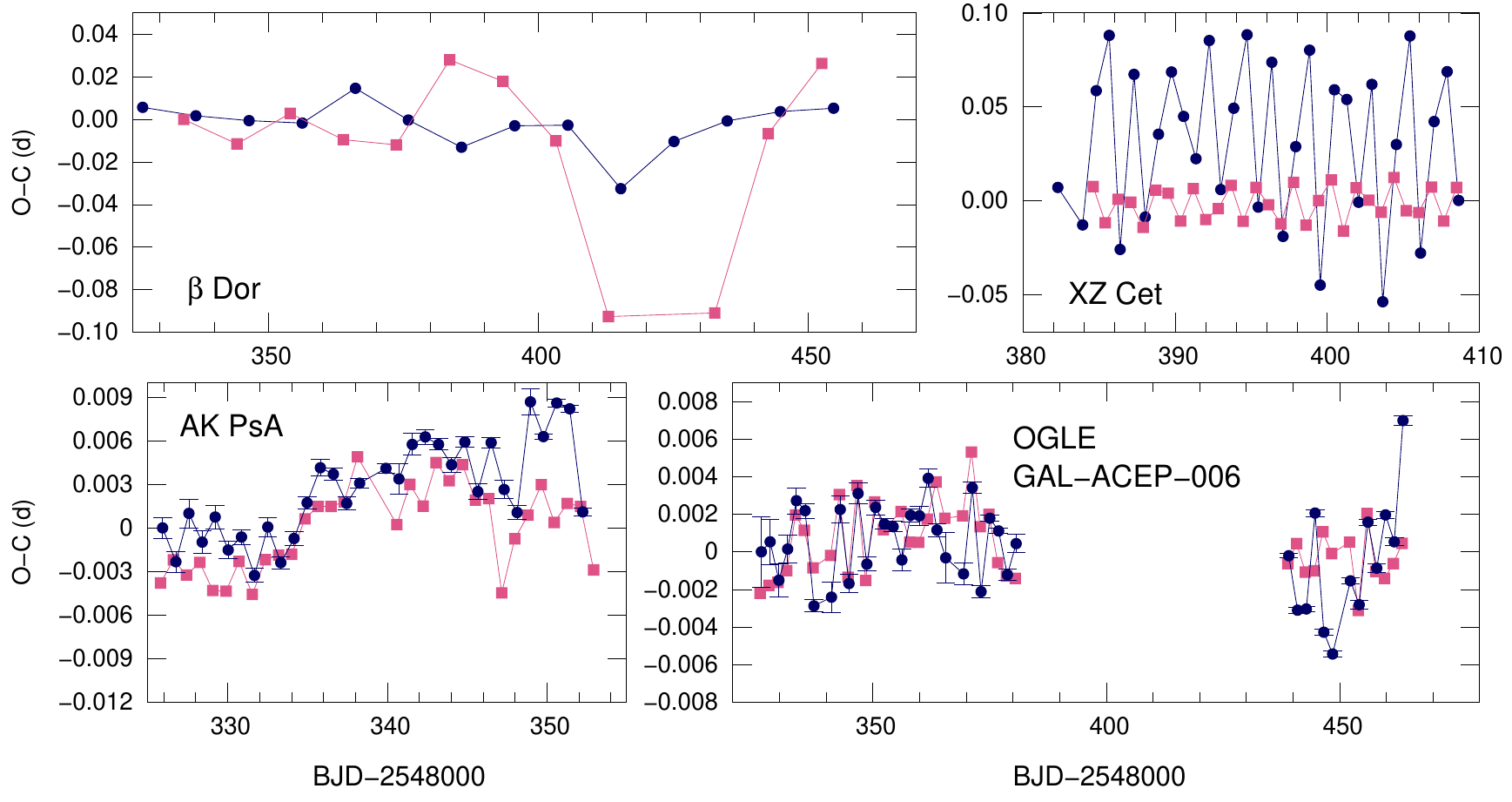}
\end{center}
\caption{O$-$C diagrams of $\beta$~Dor, XZ Cet, AK PsA, and OGLE GAL-ACEP-006, calculated via polynomial fitting of the maxima (blue) and median value of the ascending branch (pink). Errors are smaller than the symbols in the upper panels (15-20 sec for $\beta$~Dor, 1-2 sec for XZ Cet). The median method for AK PsA and OGLE GAL-ACEP-006 gives errors below 1 min.}
\label{fig:oc}
\end{figure*}

Several stars in our sample show light curve fluctuations, but in the dense stellar fields it may be impossible to unambiguously determine whether these are intrinsic features or only systematics and instrumental signals. Instead of disentangling the photometry, we can use the O$-$C technique to detect fluctuations in the time domain. We used this method to recover the light-time effect caused by a companion, as we presented in Sect.~\ref{sec:227}. In this section we discuss our results for the Galactic Cepheids. 
 
There are various methods to measure the onset and observed length of a pulsation cycle, depending on which characteristic points of the light curve are used and how they are fitted. \citet{derekas-2012} showed that the median brightness on the rising branch provides the most precise way to construct an O$-$C diagram for V1154~Cyg, the \textit{Kepler} Cepheid. For \textit{TESS}, however, flux loss and contamination may affect the moment of the median brightness, thus O$-$C values have been calculated from the timing of the sharp light curve maxima, too. 

In Fig.~\ref{fig:oc} we present O$-$C diagrams for four stars: $\beta$~Dor, XZ~Cet, AK~PsA, and OGLE~GAL-ACEP-006. We used two different methods to calculate the O$-$C values where the C values are always based on the mean pulsation period of the star. We determined the times of maxima using a third-order polynomial for $\beta$~Dor, and fifth-order polynomials for the rest of the stars to locally fit the maxima (blue points). Polynomials with different orders were tested before selecting these values. In the second method we calculated the times of the median of the ascending branch (pink). Ephemerides are given in Table~\ref{tab:eph}.

A trend is clearly visible in the O$-$C curves of AK~PsA. This O$-$C variation along with the detected amplitude modulation at the millimagnitude level (see Section \ref{sec:targets}) suggests that AK~PsA may be a Blazhko ACEP-FO, where the modulation is so weak that only precise and continuous observations from space could recover it. Since the proposed modulation cycle was not fully covered, we will need more observations to confirm this detection and determine the modulation period.

In the O$-$C diagram of the ACEP-F-type star OGLE~GAL-ACEP-006 we see strong irregular period jitter on the order of 5-10 minutes (depending on the technique) instead of a coherent trend. In comparison, most of our targets show scatter of a few minutes in their O$-$C values. It is not clear yet if the larger scatter in the pulsation period of OGLE~GAL-ACEP-006 is caused by systematics an/or elevated noise in the detectors and processing, or it is a physical fluctuation in the pulsation, which was observed in other Cepheids before. We can only answer this question after more \textit{TESS} Cepheid observations have been processed. 

The O$-$C values of $\beta$~Dor are consistent with the historical O$-$C diagram, considering the scatter present in the ground-based data \citep{berdnikov2003}. We also examined the cycle-to-cycle variation and detected slight changes, but the curve is dominated by a strong drop in the times of the median of the ascending branch in Sector 4. These data points coincide with the maxima that suffered flux loss in the 2-minute observations, indicating that even seemingly small photometric inaccuracies can cause noticeable problems during data analysis. 

Finally, the O$-$C curve of XZ~Cet clearly shows fast variations. The frequency of the variations is 0.76 d$^{-1}$, which is equal to the $f_2-f_1$ difference between the two strongest modes present in the star. The same signal is present in the median of the ascending branch but with a smaller amplitude. Clearly, the O$-$C variations here are primarily driven by mode beating instead of variations in either of the modes. Similar beating between the overtone mode and the $f_X$ mode was described in RRc stars, too \citep{hippke2015}. These findings demonstrate that intrinsic variations of the first overtone cannot be always detected via conventional O$-$C methods: in those cases temporal variations in Fourier amplitudes and phases are more reliable indicators.

Accurate O$-$C data from the TESS observations can also be used to compound existing long-term data sets to study evolutionary period changes. These in turn can then indicate in what direction are certain Cepheids crossing the instability strip \citep{turner-2006}. However, such studies require decades-long baselines and thus coverage beyond the capabilities of TESS, for now.  

\begin{table}
\renewcommand{\thetable}{\arabic{table}}
\centering
\caption{Ephemerides used to plot O$-$C graphs} 
\label{tab:eph}
\begin{tabular}{ll}
\hline
\hline
$\beta$~Dor & $C_{\mathrm{MAX}}$=$\mathrm{JD}2458326.7988$+$9.84318E$\\
& $C_{\mathrm{MED}}$=$\mathrm{JD}2458334.4241$+$9.84318E$\\
XZ Cet & $C_{\mathrm{MAX}}$=$\mathrm{JD}2445285.4260$+$0.822974E$\\
& $C_{\mathrm{MED}}$=$\mathrm{JD}2445285.2760$+$0.822974E$\\
AK PsA  &$C_{\mathrm{MAX}}$=$\mathrm{JD}2458325.9529$+$0.82210E$\\ &$C_{\mathrm{MED}}$=$\mathrm{JD}2458325.8086$+$0.82210E$\\
GAL-ACEP-006 &$C_{\mathrm{MAX}}$=$\mathrm{JD}2458326.1237$+$1.883636E$\\
&$C_{\mathrm{MED}}$=$\mathrm{JD}2458325.9206$+$1.883636E$\\
\hline 
\end{tabular}

\end{table}

\section{Summary and conclusions} 

We investigated 26 stars previously classified as Cepheids, observed in the first sectors of the \textit{TESS} mission. 
Our main goal was to explore the potential of \textit{TESS} data for two fields with different stellar densities: the Magellanic System and the Galaxy. We focused on the most intriguing phenomena that have been recently detected in Cepheids from high-quality photometric data, namely the instability of the pulsation and low-amplitude modes. We examined both short-cadence targets and full-frame image objects. For the latter we generated photometric data with the FITSH package. Our findings are as follows:

\begin{itemize}[noitemsep]
    \item [$-$] We discovered intrinsic cycle-to-cycle variations in the pulsation of the bright fundamental mode classical Cepheid, $\beta$~Dor (Fig.~\ref{fig:betadorlc}). We also demonstrated that custom-aperture photometry can result in a better light curve than the SPOC solution for this large-amplitude saturated star (Fig.~\ref{fig:betadormask}). 
    \item [$-$] We performed standard Fourier analysis and determined the main pulsation frequencies, the number of detectable harmonics, and Fourier parameters ($R_{21}$, $R_{31}$, $\phi_{21}$,  $\phi_{31}$) for all of our target stars (Table~\ref{tab:results}).
    \item [$-$] We revised classifications for two stars. We concluded RV Men is not a pulsating star, and found that OGLE SMC-CEP-4952 is more likely pulsating in the first overtone mode instead of the fundamental mode.
    \item [$-$] We discovered additional low-amplitude frequencies in an overtone anomalous Cepheid, XZ~Cet (Fig.~\ref{fig:xz}), which match the middle and lower sequences of the 0.61-type non-radial modes in the Petersen diagram of Cepheids (Fig.~\ref{fig:peter}). This is the first detection of non-radial pulsation in an AC-type star. We confirmed the stronger $f_{0.61}$ mode with ground-based multi-band measurements (Fig.~\ref{fig:hmb}), which also allowed us to determine the amplitude ratios with the overtone mode in four colours (Table~\ref{tab:xzcet_tab}).
    \item [$-$] Our analysis revealed incoherent peaks in the frequency spectra of the two Galactic ACs, suggesting instabilities in the pulsation. O$-$C diagrams confirm this assumption  (Fig.~\ref{fig:oc}). We propose the presence of period jitter in OGLE~GAL-ACEP-006 and of a low amplitude modulation in AK PsA. 
    \item [$-$] High-order harmonic series are easily detectable in the short-period Galactic Cepheids. We plotted the sequence of the Fourier amplitude ratios (Fig.~\ref{fig:harm}), and found that some stars show a sharp minimum similar to the amplitude ratio minimum previously detected in RR~Lyrae stars. The connection between the positions of the mimima and the pulsation period or mode is not yet clear from this small sample, but may be the subject of future studies with a larger number of short period Cepheids observed by \textit{TESS}. 
    \item [$-$] We investigated whether binarity can be recognised with classical O$-$C methods. For this test we chose a Cepheid in a known eclipsing binary system, OGLE~LMC-CEP-227, located far away from the most crowded stellar fields of the LMC. The O$-$C plot visibly follows the calculated light-time effect, but the large error ranges make the fit somewhat uncertain (Fig.~\ref{fig:227}). The light-time effect of OGLE~LMC-CEP-227 is quite weak due to the relatively small orbit size of the binary system. 
    More clear detection of the light-time effect can be achieved in cases where it significantly exceeds the scatter of the O$-$C technique, which strongly depends on the crowding of the stellar field, the variation of the star, and the configuration of the binary system.
    \item [$-$] Space-based photometry of peculiar W Vir type stars was analysed for the first time. We detected strong cycle-to-cycle variations that are characteristic for the W Vir type. OGLE LMC-T2CEP-023 is in an eclipsing binary system where both eclipses and ellipsoidal variation are detectable from the \textit{TESS} light curve, even though the star is strongly blended (Fig.~\ref{fig:23}).
    \item [$-$] Our investigations in the Magellanic System are confined to the outskirts (Fig.~\ref{fig:ccd}), but we faced problems from unresolved neighbouring variables even in these regions. This complication makes the identification of low-amplitude additional modes especially difficult. We presented two cases where the pulsation frequencies of another close-by star confused the Fourier analysis (Fig.~\ref{fig:4419} and \ref{fig:sxtuc}).
    \item [$-$] We detected a possible $f_{0.61}$ mode in a DCEP-1O type star in the region of the Magellanic Bridge, OGLE~LMC-CEP-3377 (Fig.~\ref{fig:peter}). This frequency is not seen in the OGLE data of the star. The low amplitude of the mode (2 mmag) can explain why it was not detected before and why no combination peaks can be found in the TESS data. However, this also means we cannot rule out a chance frequency alignment of a blended signal either.
    \item [$-$] We also discussed the possibility to estimate physical parameters of Cepheids observed with \textit{TESS} by a comparison of their precise Fourier parameters with those based on theoretical models (Fig.~\ref{fig:mass}). Modelling unprecedentedly precise \textit{TESS} light curves can potentially result in useful constraints for the pulsation models. 
    

\end{itemize}

The analysis of our Galactic field sample suggests that \textit{TESS} will result in a breakthrough in photometric  studies of short period Cepheids. The subtle light curve details can potentially reform the classification of variable stars, as well as the search for physical parameter tracers. The data of $\beta$~Dor also illustrates the importance of having long temporal baselines to study the stability of pulsation even for stars usually thought to be highly regular. Extensions to the mission of \textit{TESS} will make it possible to study the stability of Cepheids with even longer periods and/or outside the CVZs.

Considering the number of the resolved objects and the length of the observations, \textit{TESS} cannot be competitive with the OGLE Survey in the study of the Magellanic System. Despite this, we showed that \textit{TESS} is able to recover fine details in the light curves of certain stars in the outskirts of those galaxies. Some discoveries, however, will remain tentative due to the low resolution and will require independent confirmation.

The analysis of \textit{TESS} data from later sectors is beyond the scope of this paper. The whole first year of \textit{TESS} observations provides us with sufficient coverage to study the long period, bright Cepheids in the LMC and to search for amplitude and phase modulations in the short period Galactic Cepheids. Cepheids within the Galactic disk may suffer from similar confusion and blending problems that we experienced in the Magellanic System. 

Nevertheless, \textit{TESS} provides a detailed snapshot of the pulsational behaviour and offers a unique angle and valuable pieces of information that we would be unable to attain otherwise. 

\section{Acknowledgements} 

We thank the referee for the detailed comments that helped to improve the paper. The research leading to these results has been supported by the Hungarian National Research, Development and Innovation Office (NKFIH) grants KH\_18 130405, K-125015, GINOP 2.3.2-15-2016-00033 and NN-129075, and the Lend\"ulet LP2014-17 and LP2018-7/2020 grants of the Hungarian Academy of Sciences (HAS). The work reported on in this publication has been supported by the MW-Gaia COST Action (CA18104). Partial funding of the computational infrastructure and database
servers are received from the grant KEP-7/2018 of the HAS.
E.P. was supported by the J\'anos Bolyai Research Scholarship of the HAS. L.M. was supported by the Premium Postdoctoral Research Program of the HAS. C.C.N. thanks the funding from Ministry of Science and Technology (Taiwan) under the contract 107-2119-M-008-014-MY2. Funding for the Stellar Astrophysics Centre is provided by The Danish National Research Foundation (Grant agreement no.: DNRF106). R.Sm. was supported by the National Science Center, Poland, Sonata BIS project 2018/30/E/ST9/00598. M.I.J. was supported by the Ministry of Education, Science and Technological Development (contract number 451-03-68/2020-14/200002) of Republic of Serbia. A.B. acknowledges a Gruber fellowship 2020 grant sponsored by the Gruber Foundation and the International Astronomical Union and is supported by the EACOA Fellowship Program under the umbrella of the East Asia Core Observatories Association, which consists of the Academia Sinica Institute of Astronomy and Astrophysics, the National Astronomical Observatory of Japan, the Korea Astronomy and Space Science Institute, and the National Astronomical Observatories of the Chinese Academy of Sciences. J.P.G. acknowledges financial support from the State Agency for Research of the Spanish MCIU through the ``Center of Excellence Severo Ochoa'' award to the Instituto de Astrof\'isica de Andaluc\'ia (SEV-2017-0709) and from Spanish public funds for research under project ESP2017-87676-C5-5-R. P.M. acknowledges support from the NCN grant no. 2016/21/B/ST9/01126. M.S. acknowledges the financial support of the Operational Program Research, Development and Education -- Project Postdoc@MUNI (No. CZ.02.2.69/0.0/0.0/16\_027/0008360) and the MSMT Inter Transfer program LTT20015. P.K. and N.N. acknowledge the support of the French Agence Nationale de la Recherche (ANR), under grant ANR-15-CE31-0012-01 (project UnlockCepheids). 
The research leading to these results has received funding from the European Research Council (ERC) under the European Union's Horizon 2020 research and innovation program (grant agreement No. 695099).
This work has made use of data from the European Space Agency (ESA) mission {\it Gaia} (\url{http://www.cosmos.esa.int/gaia}), processed by the {\it Gaia} Data Processing and Analysis Consortium (DPAC, \url{http://www.cosmos.esa.int/web/gaia/dpac/consortium}).
Funding for the DPAC has been provided by national institutions, in particular the institutions participating in the {\it Gaia} Multilateral Agreement.
This paper includes data collected by the \textit{TESS} mission. Funding for the \textit{TESS} mission is provided by NASA’s Science Mission directorate. Resources supporting this work were provided by the NASA High-End Computing (HEC) Program through the NASA Advanced Supercomputing (NAS) Division at Ames Research Center for the production of the SPOC data products.

\software{lightkurve \citep{lightkurve1,lightkurve}, Period04 \citep{period04}, astropy \citep{astropy}, numpy \citep{numpy}, gnuplot}

\pagebreak

\appendix
\section{Photometric data tables}
Here we provide the photometric data we produced for the 30-min cadence targets with the FITSH code in Table~\ref{tab:8}, and the custom-aperture photometry of the 2-min cadence data of $\beta$~Dor in Table~\ref{tab:9}. 


\begin{table*}[ht!]

\renewcommand{\thetable}{\arabic{table}}
\centering
\caption{Sample table of FITSH photometry prepared for 23 full-frame image target stars analysed in this paper. The entire table is available online.} \label{tab:fitsh}
\begin{tabular}{llcccc}
\tablewidth{0pt}
\hline
\hline
ID & BJD(d) & Brightness(Tmag) & Brightness\_error(Tmag) & Flux(e$^-$/s) &	Flux\_error(e$^-$/s) \\
\hline
AA\_Gru & 2458325.327953 &	12.854427 &	0.001479 &	1186.394257 &	1.615479 \\
AA\_Gru & 2458325.348784 &	12.857121 &	0.001650 &	1183.454257 &	1.797872 \\
AA\_Gru & 2458325.369615 &	12.856608 &	0.001601 &	1184.014257	& 1.745759 \\
AA\_Gru & 2458325.390455 &	12.860107&	0.001990 &	1180.204257	 & 2.162657 \\
AA\_Gru & 2458325.411286 &	12.861248 &	0.001808 &	1178.964257 &	1.962894 \\
AA\_Gru & 2458325.432116 &	12.861700 &	0.002137 &	1178.474257 &	2.318994 \\
AA\_Gru & 2458325.452957 &	12.870441 &	0.002033 &	1169.024257	& 2.188713 \\
AA\_Gru & 2458325.473787 &	12.876588 &	0.002142 &	1162.424257	& 2.292938 \\
\dots & & & & & \\
\hline 
\label{tab:8}
\end{tabular}
\end{table*}


\begin{table*}[ht!]
\renewcommand{\thetable}{\arabic{table}}
\centering
\caption{Sample table of the 2-minute cadence custom aperture photometry prepared for $\beta$~Dor. The entire table is available online.} \label{tab:fitsh}
\begin{tabular}{lcccc}
\tablewidth{0pt}
\hline
\hline
 BJD(d) & Brightness(Tmag) & Brightness\_error(Tmag) & Flux(e$^-$/s) &	Flux\_error(e$^-$/s) \\
\hline
2458325.29697147 &	3.021723 &	0.000036 &	10169760 &	335.4976 \\
2458325.29836037 &	3.021729 &	0.000036 &	10169710 &	335.4959 \\
2458325.29974927 &	3.021523 &	0.000036 &	10171640 &	335.5282 \\
2458325.30113817 &	3.021536 &	0.000036 &	10171510 &	335.5254 \\
2458325.30252707 &	3.021595 &	0.000036 &	10170960 &	335.513 \\
2458325.30391597 &	3.021589 &	0.000036 &	10171020 &	335.5159 \\
2458325.30530487 &	3.021499 &	0.000036 &	10171860 &	335.5336 \\
2458325.30669377 &	3.021448 &	0.000036 &	10172340 &	335.5405 \\
2458325.30808267 &	3.021420 &	0.000036 &	10172600 &	335.5457 \\
2458325.30947156 &	3.021392 &	0.000036 &	10172860 &	335.5475 \\
\dots & & & & \\
\hline 
\label{tab:9}
\end{tabular}
\end{table*}

\end{document}